%% file: main.tex
\documentclass{article}
\usepackage[utf8]{inputenc}
\usepackage[english]{babel}
\usepackage{graphicx, color}
\usepackage[a4paper,margin=2cm]{geometry}
\usepackage{hyperref}
\usepackage[numbers]{natbib}
\usepackage{subfiles}
\usepackage{csquotes}
\usepackage{titlesec}
\usepackage{textcomp}  
\usepackage{siunitx}
\usepackage{epigraph}
\usepackage{amsfonts}
\usepackage{algorithm}
\usepackage{algpseudocode}
\usepackage{mathtools}

\titleformat{\paragraph}
{\normalfont\normalsize\bfseries}{\theparagraph}{1em}{}
\titlespacing*{\paragraph}
{0pt}{3.25ex plus 1ex minus .2ex}{1.5ex plus .2ex}

\begin{document}

\subfile{title-page-ai.tex}

\pagebreak

\epigraph{
    Truly solving program synthesis is the last programming problem mankind will have to solve.
}{
    \textit{\citet{nps}}
}


\pagebreak

\tableofcontents

\pagebreak

\begin{abstract}

    In this thesis we look into programming by example (PBE),
    which is about finding a program mapping given inputs to given outputs.
    PBE has traditionally seen a split between formal versus neural approaches,
    where formal approaches typically involve deductive techniques such as SAT solvers and types,
    while the neural approaches involve training on sample input-outputs with their corresponding program,
    typically using sequence-based machine learning techniques such as LSTMs~\citep{lstm}.
    As a result of this split, programming types had yet to be used in neural program synthesis techniques.

    We propose a way to incorporate programming types into a neural program synthesis approach for PBE.
    We introduce the Typed Neuro-Symbolic Program Synthesis (TNSPS) method based on this idea,
    and test it in the functional programming context to empirically verify type information
    may help improve generalization in neural synthesizers on limited-size datasets.

    Our TNSPS model builds upon the existing Neuro-Symbolic Program Synthesis (NSPS)~\citep{nsps},
    a tree-based neural synthesizer combining info from input-output examples plus the current program,
    by further exposing information on types of those input-output examples,
    of the grammar production rules, as well as of the hole that we wish to expand in the program.

    We further explain how we generated a dataset within our domain,
    which uses a limited subset of Haskell as the synthesis language.
    Finally we discuss several topics of interest that may help take these ideas further.
    For reproducibility, we release our code publicly.~\citep{code}

\end{abstract}

\section{Research Direction} 

If \emph{AI} is software 2.0~\citep{software20},
then \emph{program synthesis} lets us apply software 2.0 to software development itself.

\subsection{Program Synthesis}


After chess engine \emph{Deep Blue} defeated grandmaster Kasparov in 1997~\citep{deepblue},
in a freestyle chess tournament in 2005,
both a supercomputer and a grandmaster with a laptop lost to two amateurs using three laptops~\citep{kasparov},
demonstrating the importance of man-machine cooperation.

While time has passed since then, this lesson remains relevant today.
Human engineers take time to accrue experience and write software,
while even our largest generative models require human feedback for non-trivial tasks~\citep{gpt3code}.

Again, at the heart of this lies having to face the complementary strengths of people versus machines,
and in this case, applying these to improve the process of software development.
This idea of machines producing software is called \emph{program synthesis}~\citep{church1957applications}.


Historically, program synthesis has been popularized by
Microsoft Excel's \emph{FlashFill} feature~\citep{flashfill},
as well as by intelligent code completion tools,
such as Microsoft's \emph{Intellisense}~\citep{intellisense},
Google's \emph{ML Complete}~\citep{mlcomplete},
as well as Codota's \emph{TabNine}~\citep{tabnine}.

Formally speaking, program synthesis is the task of automatically constructing a program
that satisfies a given high-level specification,
be it a formal specification, a natural language description,
full program \emph{traces}, input-output examples,
or an existing program.~\citep{gulwani2017program}.


This enables us to distill our modeled program
to a simplified discrete form that may well be intelligible to humans as well as computers,
opening up opportunities for human-machine cooperation in writing software.
Specifically, this will allow machines to improve on programs written by humans, and the other way around.
As such, program synthesis may bring \emph{hybrid intelligence}~\citep{sun1994computational} to the field of software development.


For example, software engineering has brought the idea of \emph{test-driven development},
that is, the cycle of writing a test for your program to check if it does what it should,
then iterating on an implementation of the program until it passes the test.
Program synthesis may well help automate this second half.


While machine learning practicioners have gradually expanded the use-cases of AI,
GitHub in 2018 already counted 100 million code repositories~\citep{github100m},
still limited by human developers,
demonstrating the potential impact of the single AI branch of neural program synthesis.

Program synthesis itself has typically been split between formal (deductive, type-theoretic) vs. neural approaches.
This thesis aims to contribute to narrowing this gap by exploring the intersection of these approaches.

The idea of a program synthesizer utilizing a human feedback loop is not new,
having been used for ambiguity resolution when multiple programs
of different behavior both fulfilled the specified requirements.

For real-life scenarios however, the amount of viable programs might be numerous,
making it less viable to burden humans with more feedback requests than might be needed.
Neural synthesis methods such as OpenAI's GPT-3~\citep{gpt3code}
instead use sequence completion to provide the user with a likely candidate,
allowing the user to intervene as deemed fit.
This approach however has caused concern of a rise in generated code
that is neither tested nor understood.~\citep{gpt3bugs}

Synthesizer-human interaction has been further explored by the idea of \emph{type-driven development}~\citep{brady2017type},
using type annotations to inform iteratively synthesizing completions to fill program \emph{holes},
i.e. placeholder nodes in the AST to be filled by the synthesizer.
This is what directly inspired the direction of this thesis,
aiming to on the one hand facilitate such predictions where only type info by itself falls short,
while simultaneously aiming to improve on existing neural synthesis methods by also utilizing such type info.

We will now further expand on fields related to program synthesis,
before going into the challenges faced by different synthesis methods,
then lay out our research questions.

\subsection{Related fields}

To give more context on how program synthesis fits into the bigger picture,
we will briefly compare it to some other fields: program induction, supervised learning, as well as constraint satisfaction and discrete optimization.

\subsubsection{Program Induction}

Unfortunately the field of program induction suffers from competing definitions, blurring the distinction between what constitutes program \emph{synthesis} versus what constitutes program \emph{induction}. In short though, those in either field claim to be more general than the other branch.

The field of \emph{inductive programming}~\citep{popplestone1969experiment,plotkin1970note,fogel1966intelligent},
primarily known for its sub-branch \emph{inductive logic programming}~\citep{muggleton1991inductive} focused on propositional logic,
is simply automatic synthesis using inductive logic,
and was coined to distinguish itself from the \emph{deductive} techniques used in \citet{church1957applications}'s synthesis of circuits.
Under this definition, the term program synthesis is used to refer to its original scope of program generation using deductive techniques.

Whereas in the original problem definition the desired behavior was fully specified, program \emph{induction} aimed to generalize the problem to also tackle automatic generation of programs for which the desired behavior had only partially been made explicit, through e.g. input/output examples or incomplete data.

Under this definition, there is no significant distinction between our present work and program induction's sub-branch of \emph{inductive functional programming}, focused on the generation of programs in functional programming languages such as Lisp~\citep{lisp} or Haskell.

Nevertheless, in current parlance program \emph{synthesis} is often used in a broader scope, extending from the original deductive approach to include inductive approaches as well.
In this view, the two fields are distinguished in that program \emph{synthesis} is defined as to explicitly return a program,
whereas program \emph{induction} learns to \emph{mimic} it rather than explicitly return it.%
~\citep{devlin2017robustfill,gulwani2017program,nps}
While this usage appears to clash with the term program induction as used in inductive \emph{functional} programming,
this view of program induction being limited to this smaller scope likely stems from widespread use of the term in the field of inductive \emph{logic} programming.

This terminology itself is not of much concern for our present paper, as the boundaries between the fields have often been muddy.
Moreover, recent applications of AI to this field have led to the more recent term of \emph{neural program synthesis}~\citep{nps}. Therefore, we will simply settle for using `program synthesis' to refer to the field in general as well.

\subsubsection{Supervised Learning}

The above definition of program synthesis as explicitly returning a program
is helpful to explain how it differs from \emph{supervised learning},
the machine learning task of learning a function that maps an input to an output based on example input-output pairs~\citep{russell2002artificial}.

Deep learning methods may potentially be applied to different branches of program synthesis,
and several of these may in fact be tackled using setups involving supervised learning.
Of particular note here however is a branch referred to as \emph{programming by example} (PBE),
which like supervised learning is based on the question of how to reconstruct a mapping between input and output --- in the supervised learning context also referred to as \emph{features} and \emph{labels}, respectively.

What sets these apart is that,
whereas supervised learning would construct such a model in a continuous vector space,
allowing probabilistic interpretations of the data to be taken at prediction time,
PBE instead fits its model into the discrete form of a given \emph{grammar} to produce a program,
forcing one to instantiate such a model from probabilistic data interpretations.

This also explains the relative benefits of these two fields:
supervised learning enables differentiable evaluation
without needing workarounds (see Section \ref{sec:neuralpbe}),
allowing for optimization by backpropagation~\citep{backproprnn},
and is not limited in expressivity by the limitations of any particular grammar or set of operations.
This makes it well-positioned to solve problems deemed too complex for traditional programming, such as image recognition.

Program synthesis techniques may instead construct a traditional program,
which
can generalize better~\citep{nps}, be provably correct~\citep{nps},
as well as potentially faster to execute than predictions using the equivalent supervised learning model.

Moreover, using programs as a common denominator between human and machine-based programmers makes for human-intelligible machine-made models,
relevant in the field of \emph{interpretable} or \emph{explainable artificial intelligence}
~\footnote{
    If the goal in using program synthesis is to make models more interpretable,
    one could potentially start out by training a neural model,
    then approximate this by synthesizing a program similar to it.
    And in fact, \citet{pirl} apply exactly this approach for reinforcement learning.
},
while also enabling human-machine cooperation in the production and maintenance of software.


In program synthesis, one may take an existing program, and synthesize variants intended to generalize the existing logic to match the new data.~\citep{myth}
This makes program synthesis well-suited to facilitate the automation of programming.%
~\footnote{
    One may note that this would technically enable the synthesis of programs implementing machine learning models as well.
    However, such an approach would make for a relatively expensive evaluation function,
    and as such is traditionally left to the field of \emph{neural architecture search}.
}

In other words, whereas supervised learning makes for simpler learning,
since it foregoes the need to define a synthesis grammar and operator set,
program synthesis may make for programs that are potentially more efficient,
more understandable,
which for the machine learning models produced in supervised learning requires adding the non-trivial field of \emph{explainable AI}~\citep{gunning2017explainable},
while program synthesis also facilitates incorporating knowledge of human experts,
by allowing them to offer relevant operators.

\subsubsection{Constraint satisfaction vs. discrete optimization}

While its definition may appear to frame program synthesis as a type of \emph{constraint satisfaction} problem (CSP),
where a program either does or does not satisfy the given specification,
one could also opt to approach it as a \emph{discrete optimization} problem,
as specifications such as input-output examples allow us to count the examples our candidate program satisfies.

Intuitively, a program satisfying part of our examples may be regarded as closer to a solution than one that does not satisfy as many.
Furthermore, additional considerations such as performance may further push us to find a solution that not only calculates outputs correctly but also runs within reasonable time or memory constraints.
These then provide a quantifiable feedback measure for us to optimize.

However, constraint satisfaction and discrete optimization were intended to solve fully specified problems (\emph{deduction}),
while in modern-day program synthesis, as we will explain later,
we usually need to settle for a \emph{partial} specification of the intended program's behavior (\emph{induction}).

In other words, in such an inductive setting one cannot even definitively tell whether one program is better or worse than another on unspecified \emph{intended} behavior,
meaning the metric that would be used for constraint satisfaction or optimization may not be representative of the actual problem.
This is a general difference between constraint satisfaction and optimization versus pattern recognition techniques including machine learning:
in the latter case, the actual goal is to \emph{generalize} learned behavior to an unseen test set,
rather than merely performing well on known examples.

As such, applying such techniques to the field of program synthesis using partial specifications may lead us to the problem of \emph{overfitting}:
while found solutions might well satisfy the \emph{specified} behavior,
the question would be whether these would also generalize to match our \emph{intended} behavior,
as is the goal in PBE.

\subsection{Challenges}

Having given a brief background on where the field of program synthesis fits in,
we will now briefly outline some of the challenges in this field,
with a focus on programming by example,
as a backdrop informing our own research direction.

\subsubsection{Challenge of program synthesis}

While considered a holy grail of computer science~\citep{gulwani2017program},
program synthesis in general is a challenging task, characterized by large search spaces,
e.g. a search space of $10^{5,943}$ programs to discover an expert implementation of the MD5 hash function.~\cite{gulwani2017program}

\subsubsection{Challenge of type-theoretic programming by example} \label{sec:typepbe}

One issue with type-theoretic approaches to PBE, later introduced in further detail,
is that while such search methods are able to make use of both input/output examples and types in their search,
there is no sense of learning across problem instances to further reduce the synthesis time caused by these large search spaces.

\subsubsection{Challenges of neural programming by example} \label{sec:challengesnps}

For neural methods in PBE, the original challenge of large search spaces means
it will not be viable to proportionally scale our training sets by program size.

Furthermore, whereas a program synthesizer may be programmed or taught to output programs adhering to a given grammar,
we may generally only be able to evaluate the quality of \emph{complete} programs:
there is typically no guarantee that \emph{partial} constructions of the program would \emph{also} qualify as a full executable program adherent to the grammar.
As a result, neural synthesizers will have little intermediary feedback to go by, limiting their effectiveness.

But if only \emph{complete} programs can be evaluated for validity and behavior, then 
we will be ill-equipped to provide synthesizers with an accurate understanding of partial programs,
which make up for a large part of our prediction steps.
As such, it would be desirable to somehow better supervise the intermediate prediction steps.
This echoes \citet{nps}'s conclusion that one area of research in neural program synthesis that requires further exploration is
\emph{specifically designing neural architectures} to excel at the difficult problems of program synthesis.


Most neural synthesis techniques, particular those using a \emph{sequence-to-sequence} approach,
additionally face the issue of dissonance between their representation of complete programs and that of intermediate states.
As such intermediate states do not in general constitute valid programs,
these neural synthesizers have an additional task to solve:
compensating for their lack of an inherently meaningful incremental state.

\subsection{Research question}


Based on the previous section,
our key observation here is thus that input-output examples and types are
quite complementary as specifications constraining our program behavior.
Input-output examples are relatively expressive,
but may only help us to evaluate the quality of complete programs.
Types, on the other hand, are by themselves not usually descriptive enough of our task,
but may help us to provide a less noisy summary of program behavior, hopefully aiding generalization,
as well as to evaluate even incomplete programs still containing holes,
and to inform further incremental synthesis steps.

This brings us to the question:
        can neural program synthesis methods benefit from using type information?

\subsubsection{Hypothesis}

Based on the complementary strengths mentioned above,
we therefore hypothesize that program synthesizers may capitalize on this synergy by utilizing both kinds of information,
rather than settling for only one of the two, as most existing methods have done.%
~\footnote{
    While one might wonder if this constrains our idea to the subset of PBE problems where type information is available,
    this limitation is essentially meaningless:
    when one has input-output examples in a programming language supporting type inference,
    one would already have the types of these input-output examples.
    This would render our idea applicable for practically any (neural) methods for PBE.
}%

Specifically, we formulate the following hypothesis:
\begin{displayquote} 
    \emph{
        Hypothesis:
        the effectiveness of neural program synthesis may be improved by
        adding type information as additional features.
    }
\end{displayquote}




\section{Expected Contribution} 

The present work aims to be the first experiment to:
\begin{itemize}
    \item bring the type-based information traditionally used in functional program synthesis into the newer branch of neural program synthesis,
    better constraining the search space to improve the effectiveness of neural program synthesis methods;
    \item show that the neural synthesis of statically typable programs may benefit from techniques \emph{specific} to this domain, and therefore for the purpose of automatic programming merits further study in itself;
    \item offer an open-source implementation of the algorithm described in \citet{nsps};
    \item generate a dataset for neural synthesis of functional programs, and lay out how to do this, including an open-source implementation, addressing the current reliance on hand-crafted curricula~\citep{nps}.
\end{itemize}

\pagebreak

\section{Literature review} \label{sec:litreview}

To provide some background to our hypothesis,
we will use this section to first give a brief overview of how programming by example (PBE) fits into the broader picture of program synthesis,
as well as what existing approaches there have been to PBE,
including the \emph{neuro-symbolic program synthesis} model we build upon ourselves.

On types of synthesizers,
\citet{gulwani2017program} introduce a taxonomy based on three key dimensions:
\begin{itemize}
    \item the kind of \emph{constraints} that it accepts as expression of \emph{user intent};
    \item the \emph{space} of programs over which it searches;
    \item the \emph{search technique} it employs, i.e. the synthesizer.
\end{itemize}

User intent in program synthesis can be expressed in various forms, including logical specification~\citep{temporalstreamlogic} (among which types~\citep{synquid}),
input-output examples, traces, natural language~\citep{abstractsyntaxnetworks},
partial programs, or even related programs.~\citep{gulwani2017program}

While the common thread in program synthesis is that our intended output takes the form of a program,
sub-branches of this field are primarily defined by the types of input we use to come to this output,
i.e. the 
first point
in the above classification.

We will briefly describe such variants of program synthesis in the next section,
with some minor focus on search technique as influenced by this problem description.
Search technique we will explore in further depth for PBE in Section \ref{sec:pbe}.

\subsection{Types of program synthesis} \label{sec:synthtypes}


Program synthesis was traditionally studied as a computer science problem,
where the problem was typically framed using a \emph{formal specification}.
This problem was then tackled using e.g. an enumerative search, deductive methods, or constraint-solving techniques.~\citep{gulwani2017program}
However, such formal specifications ended up about as hard to write as the original program,
rendering this approach to the problem not very useful.

Closely related to this field is the idea of synthesizing a program solely from its \emph{type signature}.~\citep{djinn,synquid}
Traditionally types would make for \emph{inductive} synthesis,
i.e. only making for an \emph{incomplete} program specification,
but this may end up not sufficiently expressive:
while certainly constraining the program space,
input/output examples may still be needed to disambiguate between potential candidate programs.
Adding such examples brings us to the branch of \emph{type-theoretic} PBE,
which we will introduce in further detail in Section \ref{sec:deductivepbe}.
Program synthesis approaches using types have commonly focused on using functional programming languages as the synthesis language.%
~\citep{synquid,eguchi2018automated,scythe,scout,gissurarson2018suggesting,idris,lenses}

There have also been attempts to get such a type-based approach to become closer to deductive synthesis,
i.e. making for a complete behavioral program specification,
through the use of e.g. the more expressive refinement types~\citep{synquid} or succinct types~\citep{guospeeding}.
However, these approaches tend to fall into a similar pitfall as synthesis from \emph{formal specifications},
requiring the user to write such a detailed type specification that they might have as well just written the program directly.%
~\footnote{
    However, perhaps one might instead also be able to synthesize this detailed type specification,
    giving the benefit of additional formal guarantees from our actual program that we could then synthesize from this type specification!
}

Compared to formal specifications, it was found that for users,
input-output examples were a more attractive way to specify desired program behavior
.~\citep{bodik2013algorithmic}
This field is named \emph{programming by example} (PBE).
As the specification is incomplete here, PBE is considered \emph{inductive} synthesis, as opposed to the \emph{deductive} synthesis where we do have a complete specification.
In other words, from the perspective of the synthesizer, PBE is generally a more difficult problem.

PBE may be further split up according to the type of program to be synthesized~\citep{bodik2013algorithmic},
generating \emph{logic programs} (assigning truth values to variables),
or generating \emph{functional programs} (e.g. Lisp, Haskell).
PBE too has branches based on deductive techniques (including type-theoretic PBE),
inspired by synthesis from formal specifications.
Our work will focus on PBE in the category of functional programs,
where the goal is to automate traditional programming tasks.


Synthesis from program \emph{traces}~\citep{koskimies1994automatic},
and the related synthesis from \emph{Linear Temporal Logic} (LTL)
specifications~\citep{camacho2019towards}, are about
a system reacting to sequences of inputs to mimic the desired program behavior.
These are useful for e.g. specifying the expected behavior of user interfaces.
Essentially this task may be viewed as a generalized version of PBE,
adding the additional challenge of figuring out which inputs triggered which state changes.



\subsection{Existing approaches to programming by example} \label{sec:pbe}

PBE has traditionally known heuristics such as
\emph{Version Space Algebras} (VSAs)~\citep{mitchell1982generalization},
which aim to constrain grammar productions by using
\emph{candidate elimination} to keep track of a \emph{hypothesis space}.
Another useful tool is \emph{ambiguity resolution},
i.e. requesting user input to resolve ambiguity
in the event that multiple candidate programs
fulfill the given input-output example pairs.%
~\citep{gulwani2017program}
However,
these two techniques are primarily used to complement
other methods we will introduce here now.

It must be noted that program synthesis has been somewhat different
from other branches of machine learning, such as image recognition:
although there have been competitions like the
\emph{Syntax-Guided Synthesis competition} (SyGuS-Comp)~\citep{sygus},
unfortunately the field has been so diverse that there has been only limited
standardization of benchmarking tasks to compare approaches,
as \emph{ImageNet}~\citep{deng2009imagenet} had done for computer vision tasks.%

While this means we will not present statistics
comparing the performance of these various approaches,
we will instead lay out their conceptual differences and weaknesses.

\subsubsection{Search-based programming by example}

Under search-based methods for programming by example we will classify any approaches that do not involve \emph{learning} to synthesize by means of a neural component.
While the approaches in this category range from naive to sophisticated,
they unfortunately share a common drawback:
whereas neural synthesizers allow one to tweak a loss function to take into account various sub-goals,
non-neural synthesizers have no sense of \emph{learning}
from existing programs or across problem instances,
meaning they will have trouble achieving:
\begin{itemize}
    \item \emph{generalizability}~\citep{nps};
    \item \emph{interpretability} to humans (human \emph{source code bias}, i.e. make generated code more similar to the way it is written by humans)~\citep{nps};
    \item \emph{synthesized program performance} (as measured in e.g. raw CPU time)~\citep{schkufza2016stochastic};
    \item an increase in \emph{synthesizer performance},
    as they must solve any new synthesis task essentially from scratch,
    and could never have as much information to this end as a neural synthesizer,
    which may in fact be able to use arbitrary learned features~\citep{odena2020learning}.
\end{itemize}

\paragraph{Enumerative search}

The naive approach to synthesis would be to enumerate all the possible programs in our search space,
and for each one evaluate whether it satisfies our task specification.
This is called \emph{enumerative} or \emph{depth-first search} (DFS).
As one might expect, such an approach does not generally scale well with search space size however.

\paragraph{Oracle-guided synthesis}

One attempt to overcome the computational complexity of program synthesis
has been \emph{oracle-guided synthesis}~\citep{solar2008program},
which splits the synthesis task into generating and filling of program
\emph{sketches}~\citep{murali2017neural}.
Unlike full synthesis itself, sketch filling is not a second-order but a
first-order logic problem~\citep{gulwani2017program}, enabling the use of constraint-solving methods
such as satisfiability (SAT) or satisfiability modulo theories (SMT) solvers
(which combine SAT-style search with theories like arithmetic and inequalities)%
~\citep{akiba2013calibrating,alur2013syntax,alur2016sygus,rosette,architecture}
to fill sketches, potentially further extended with
\emph{conflict-driven learning}~\citep{feng2018program,hornclauses},
which helps \emph{backtrack} if the branch explored turns out unviable.
The point here is that if a given sketch has multiple holes,
once a filled version turns out unviable due to a certain production rule used for one of its holes,
other variants involving the faulty choice in question may be ruled out as well.

This synthesis method has also spawned a solver-aided language
designed to facilitate this type of program synthesis~\citep{rosette},
which generates satisfiability conditions for satisfactory programs
based on failing input-output examples such as to synthesize program repairs.


\paragraph{Deductive techniques} \label{sec:deductivepbe}

\emph{Deductive} search techniques for PBE were inspired by
techniques used in synthesis from formal specifications,
but have been applied to the \emph{inductive} task of PBE as well.
Deductive techniques are based on theorem provers,
and recursively reduce the synthesis problem into sub-problems,
propagating constraints.
These include approaches based on \emph{inverse semantics} of
domain-specific language (DSL) operators and \emph{type-theoretic PBE}.


The idea of \emph{inverse semantics} is to reduce the complexity
of the synthesis task by using inverse logic.~\citep{flashmeta,prose}
This is a top-down search where we would take a grammatical
production rule, presume it to be our outer expression,
and use its inverse logic to propagate our original
input-output examples to its sub-expressions.
This way we have obtained a simpler sub-problem to solve.


While this is a useful search technique however, its
use is unfortunately limited to invertible operations,
rendering this a helpful complement to, yet not a
reliable alternative to other PBE methods.%
~\footnote{
    A recent potential workaround not reliant on invertability
    has been the approach by \citet{odena2020learning},
    who would, given properties of a function composition
    $f \circ g$ and of $f$, use machine learning to predict
    the properties of $g$.
    However, it is not immediately clear if this technique
    has a straight-forward equivalent in the domain of
    input-output examples.
}


\emph{Type-theoretic} deductive search is about the use of programming
types to constrain the synthesis search space.


While in Section \ref{sec:synthtypes} we noted such purely type-based
approaches fell into the pitfall of requiring the user to write a
type specification similar in complexity to the actual program itself,
this branch is nevertheless useful in combination with other methods,
and the use of type-theoretic deductive search has been combined with PBE.~\citep{myth}


\subsubsection{Neural programming by example} \label{sec:neuralpbe}

More recently, PBE has been explored using machine learning approaches as part of \emph{neural program synthesis}.~\citep{nps}
Whereas traditional approaches in program synthesis (and particularly PBE) focused on constraining the large discrete search space,
such as deductive and constraint-solving approaches,
\emph{neural} program synthesis generally uses \emph{autoregressive}~\citep{kendall1944autoregressive} methods,
i.e. incrementally generating programs with each prediction step depending on the previous prediction result.
Neural synthesis models use continuous representations of the state space to predict the next token,
be it in a sequential fashion~\citep{npi,neuralmachinetranslation,alphanpi},
or in a structured one based on ASTs~\citep{nsps}.

Unfortunately though, program synthesis in its general sense has been less straight-forward to tackle by neural methods than some other AI problems,
as like in \emph{natural language processing} (NLP),
our search space is typically discrete, meaning we cannot simply apply gradient-based optimization such as \emph{stochastic gradient descent} (SGD).~\citep{nps}

The issue here is that, in order to learn the parameters of a neural network, SGD uses the gradients available in a continuous search space to evaluate in which direction to adjust its parameters.
However, our program synthesis setting does not have an inherent continuous space:
it does not make sense to ask e.g. what program is half-way in between $x+x$ and $x \cdot x$.

As such, in discrete settings we lack this required notion of gradients:
while we might evaluate the quality of different programs,
we may not have intermediate programs to evaluate the quality of a given optimization direction.

This problem can be worked around in different ways:
\begin{itemize}
    \item Using a \emph{differentiable interpreter} to directly enable gradient-based optimization.~\citep{forth,terpret,houdini,feser2016differentiable,rocktaschel2017end,abadi2019simple}
        However, while only empirical evidence is available to compare this approach, as per \citet{terpret} such purely SGD-based methods so far appear to have proven less effective than traditional or mixed methods such as linear programming, Sketch~\citep{solar2008program} or SMT.
    \item Using \emph{strong supervision}, i.e. create a differentiable loss signal
        to supervise synthesis training by checking if the synthesized program is identical to the target program,
        rather than if it has equivalent behavior.
        This approach unfortunately simplifies our problem too much%
        ~\footnote{
            In reality, we wish to condition our model to synthesize not just the known programs,
            but to generalize to learn to synthesize unknown programs matching our task specification as well.
            Supervising by a given `known correct' program instead tells our model that other programs matching our specification somehow do not qualify as correct.

            As a result, such supervision requires that the training dataset provides a representative sample of our full program space:
            training on the full program search space ensures that such bias from individual samples should be approximately averaged out.
            This assumption is broken however for datasets much smaller than the program space,
            meaning that this approach does not scale well to bigger search spaces.~\citep{nsps}
        }, but does make for a relatively simple setup.
    \item Using \emph{weak supervision}~\citep{mapo},
        which tends to address the problem of reward differentiability by using \emph{reinforcement learning} techniques to estimate a gradient to optimize by%
        ~\citep{chen2017towards,bunel2018leveraging,xu2019neural,camacho2019towards},
        so as to learn to synthesize by trying based on program performance rather than from direct supervision signals.
        This approach solves the issues of supervised neural synthesis,
        but requires a more complex setup.
        This typically involves bootstrapping on strong supervision to overcome the \emph{cold start} problem of finding an initial reward gradient.
    \item Using neural methods in a \emph{hybrid} setup. This approach is explored further in Section \ref{sec:ngs}.
\end{itemize}

\paragraph{Sequence-based neural program synthesis} 

Neural synthesis methods typically employ \emph{sequence-to-sequence}
(or simply \emph{seq2seq}) techniques~\citep{npi,neuralmachinetranslation,alphanpi},
such as the \emph{recurrent neural network} (RNN)~\citep{backproprnn}
and \emph{long short-term memory} (LSTM)~\citep{lstm},
leveraging techniques commonly used in NLP
to represent program synthesis as a sequence prediction problem.%

Such sequential neural synthesizers have been extended with mechanisms such as
convolutional recurrence~\citep{neuralgpu},
attention~\citep{nmt,ptrnets,structuredattention},
memory~\citep{ntm,neuralram,neuralprogrammer,hierarchicalmemory},
function hierarchies~\citep{npi,npl},
and recursion~\citep{cai2017making}.

However, while a hypothetical synthesizer only producing compilable programs would always have direct feedback to its program embeddings,
this feedback signal is much delayed if a synthesizer would gradually synthesize a program e.g. one character at a time,
only learning about resulting program behavior once the program is complete.

As such, sequence-based neural techniques must learn quite a lot:
in addition to (continuous logical equivalents of) the traditional compiler tasks of \emph{lexing} input into token categories,
\emph{parsing} these token sequences into hierarchical structures (ASTs),
and interpreting these to execute them as programs,
these synthesizers must additionally learn how to construct and update a (memorized) state so as to ultimately,
when the synthesizer considers its code complete, obtain a correct program.

In addition, for our purposes, in sequence-based neural synthesis techniques,
any given intermediate prediction does \emph{not} necessarily \emph{itself} qualify as a program in the grammar,
meaning we are not able to apply a type-based analysis to gain further info for use in further synthesis steps.

\paragraph{Tree-based neural program synthesis}

More recently, there have also been approaches framing program synthesis by representing programs as ASTs rather than as sequences~\citep{polosukhin2018neural},
allowing such methods to use \emph{tree-structured networks}%
.

While
we previously
mentioned the issue of
sequence-based neural methods lacking an inherently meaningful incremental state,
tree-based methods should at least result in an (incomplete) \emph{abstract syntax tree} (AST).
This is significantly easier to learn to embed given the knowledge of how to embed a complete AST than it would be to embed a program that does not even parse,
as the unfilled dummy nodes or \emph{holes} may simply be added as an additional AST symbol to embed.

Of particular interest to us in this category has been the work of \citet{nsps},
which we will introduce in more detail in the next section.

\paragraph{Neuro-symbolic program synthesis} \label{sec:nsps}

The \emph{neuro-symbolic} program synthesis (NSPS) model introduced in \citet{nsps}
is the model we will build upon for our own experiment, so we will explain it in more detail here.
The reason we picked NSPS as our benchmark algorithm in particular is
that there have been only few neural synthesizers out there
based on abstract syntax trees (ASTs) rather than sequences.

NSPS is named after the fact that it uses programming \emph{symbols} as neural features,
allowing it to combine symbolic and neural approaches.
NSPS improves on existing \emph{sequence-to-sequence}-based neural synthesis models
by using a tree-based neural architecture they call the
\emph{recursive-reverse-recursive neural network} (R3NN).

\begin{figure*}[h]
    \begin{tabular}{c|c}
        \begin{minipage}{0.5\linewidth}
            \includegraphics[scale=0.3]{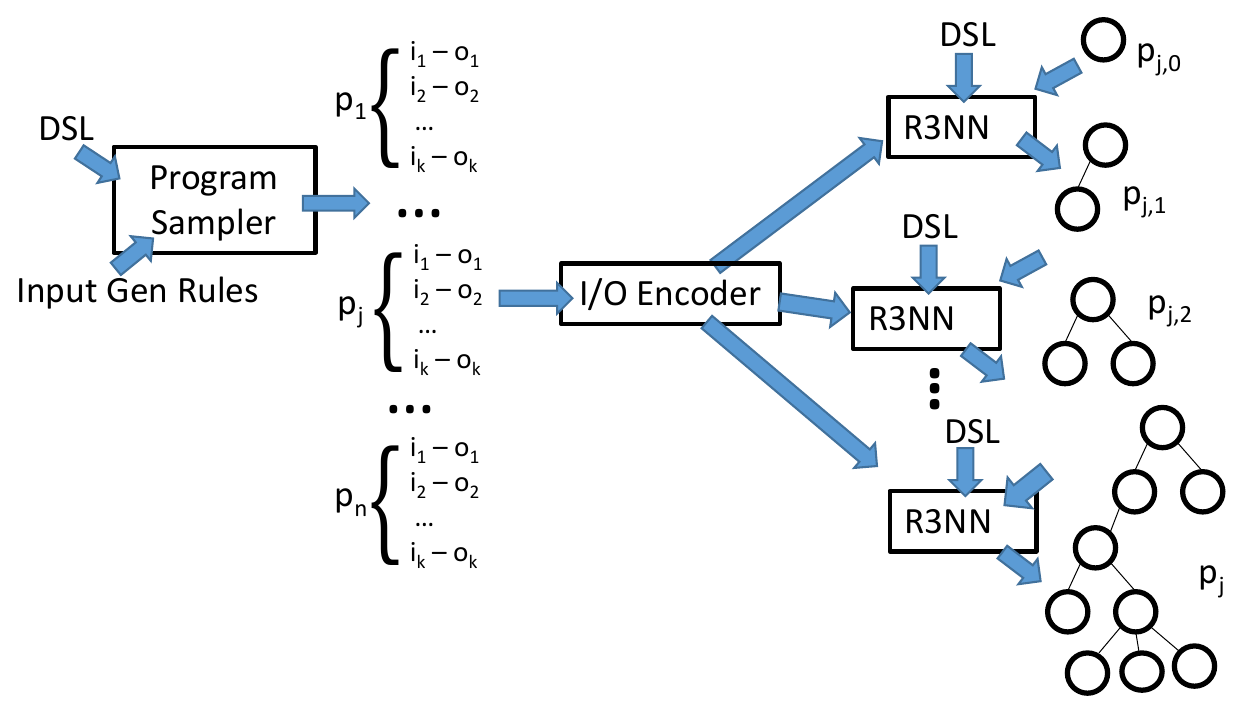}
        \end{minipage}
        &
        \begin{minipage}{0.5\linewidth}
            \includegraphics[scale=0.3]{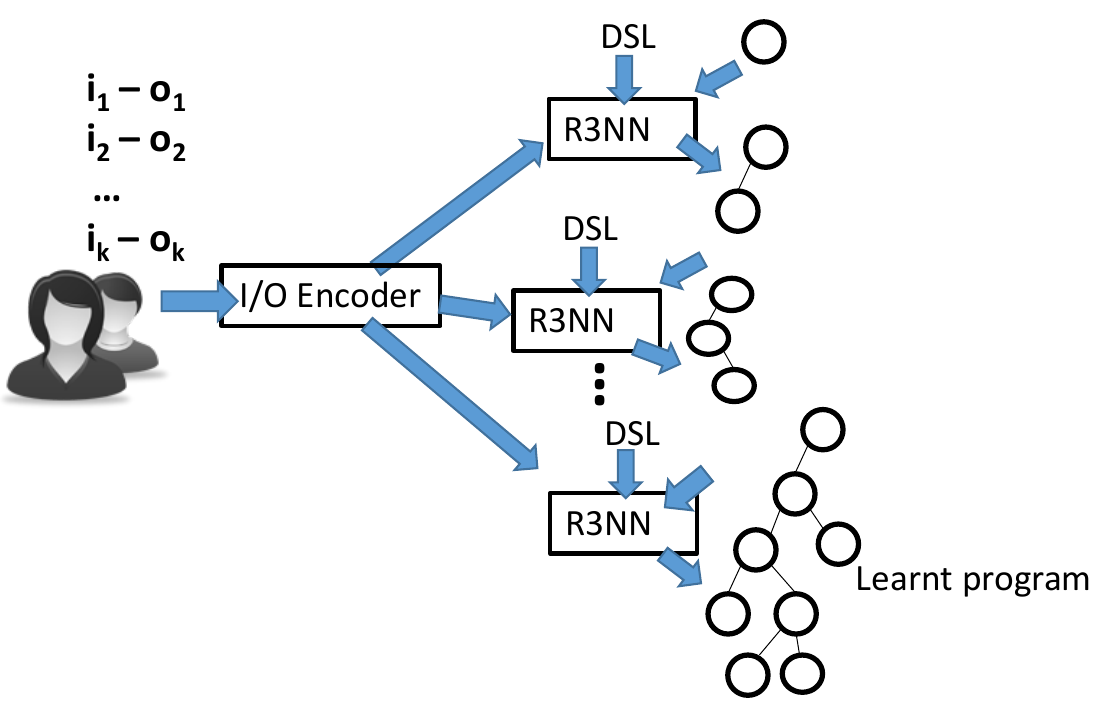}
        \end{minipage}
        \\
        (a) Training Phase & (b) Test Phase
    \end{tabular}
    \caption{overview of the Neuro-Symbolic Program Synthesis model, taken from \citep{nsps}}
    \label{nsps}
\end{figure*}

NSPS then aims to make predictions on credible rule expansions to fill holes
in \emph{partial program trees} (\emph{PPTs}) --- basically ASTs containing holes --- based on the program's content and structure.
As usual in neural PBE NSPS also conditions on the (encoded) input/output examples, as seen in Figure \ref{nsps}.
These hole expansions are based on a \emph{context-free grammar} describing the domain-specific language (DSL) to be synthesized.
Such a grammar consists of sets of expansions rules from left-hand symbols to productions in the grammar (which may include further left-hand symbols).

\citet{nsps} try out different example encoders,
each embedding into a continuous space a \emph{one-hot} representation of the input or output strings of their domain,
i.e. for a vocabulary of `a', `b' and `c', encode `b' as $010$,
meaning the second option out of three.
They start out with a simple LSTM baseline,
then introduce different variants based on the \emph{cross-correlation}~\citep{bracewell1986fourier} between inputs and outputs.

The baseline sample encoder processes input/output strings of example pairs
using two separate deep bidirectional LSTM networks,
one for inputs, one for outputs.
For each I/O pair, it then concatenates the topmost hidden representation
at every time step to produce a $4HT$-dimensional feature vector per I/O pair,
where $T$ is the maximum string length for any input or output string,
and $H$ is the topmost LSTM hidden dimension controlling the amount
of features we would like per one-hot encoded characters.
It then concatenates the encoding vectors across all I/O pairs
to get a vector representation of the entire I/O set.~\citep{nsps}


\begin{figure*}[h]
    \begin{tabular}{cc}
        \begin{minipage}{0.45\linewidth}
            \includegraphics[scale=0.16]{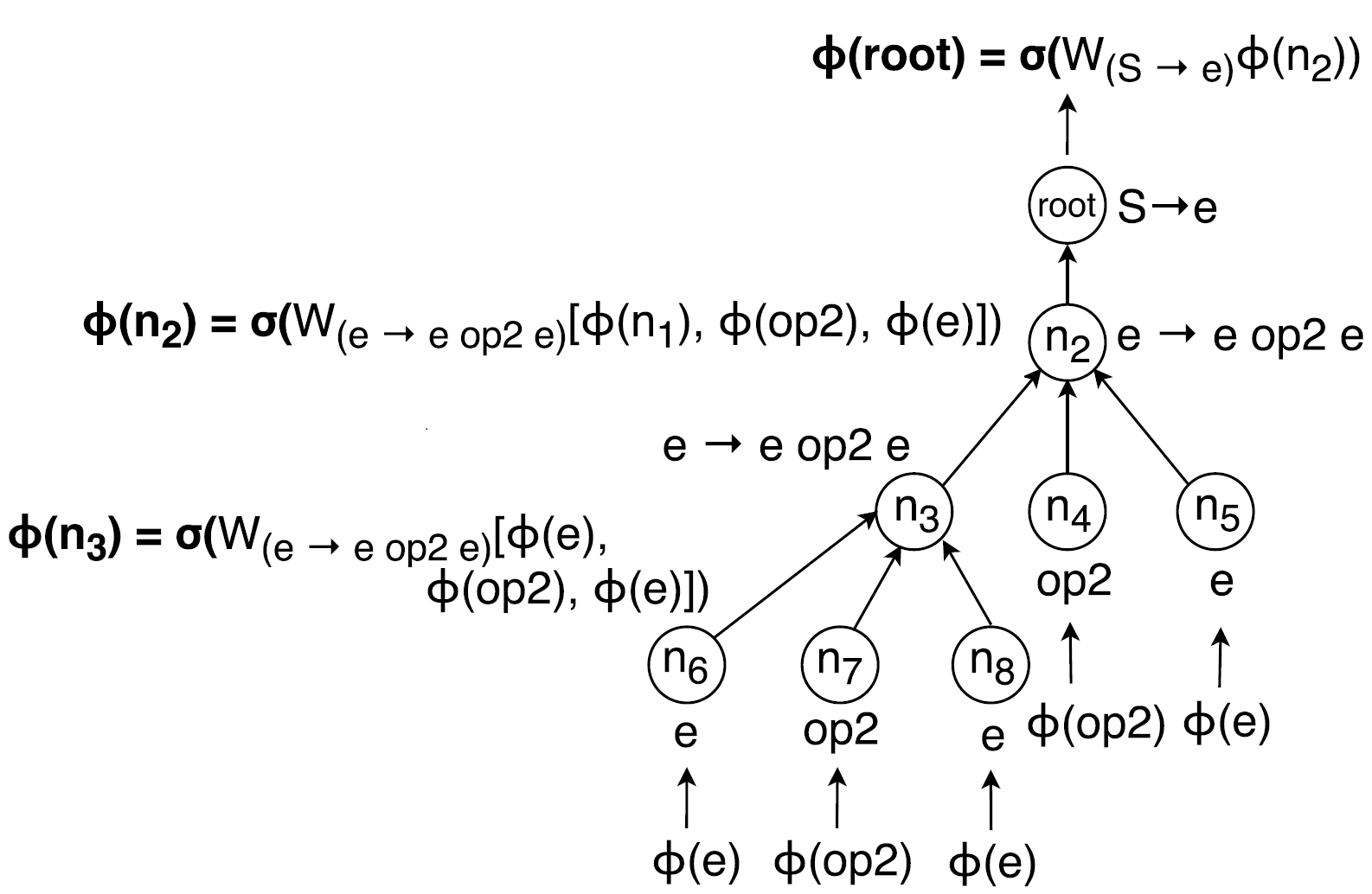}
        \end{minipage}
        &
        \begin{minipage}{0.55\linewidth}
            \includegraphics[scale=0.16]{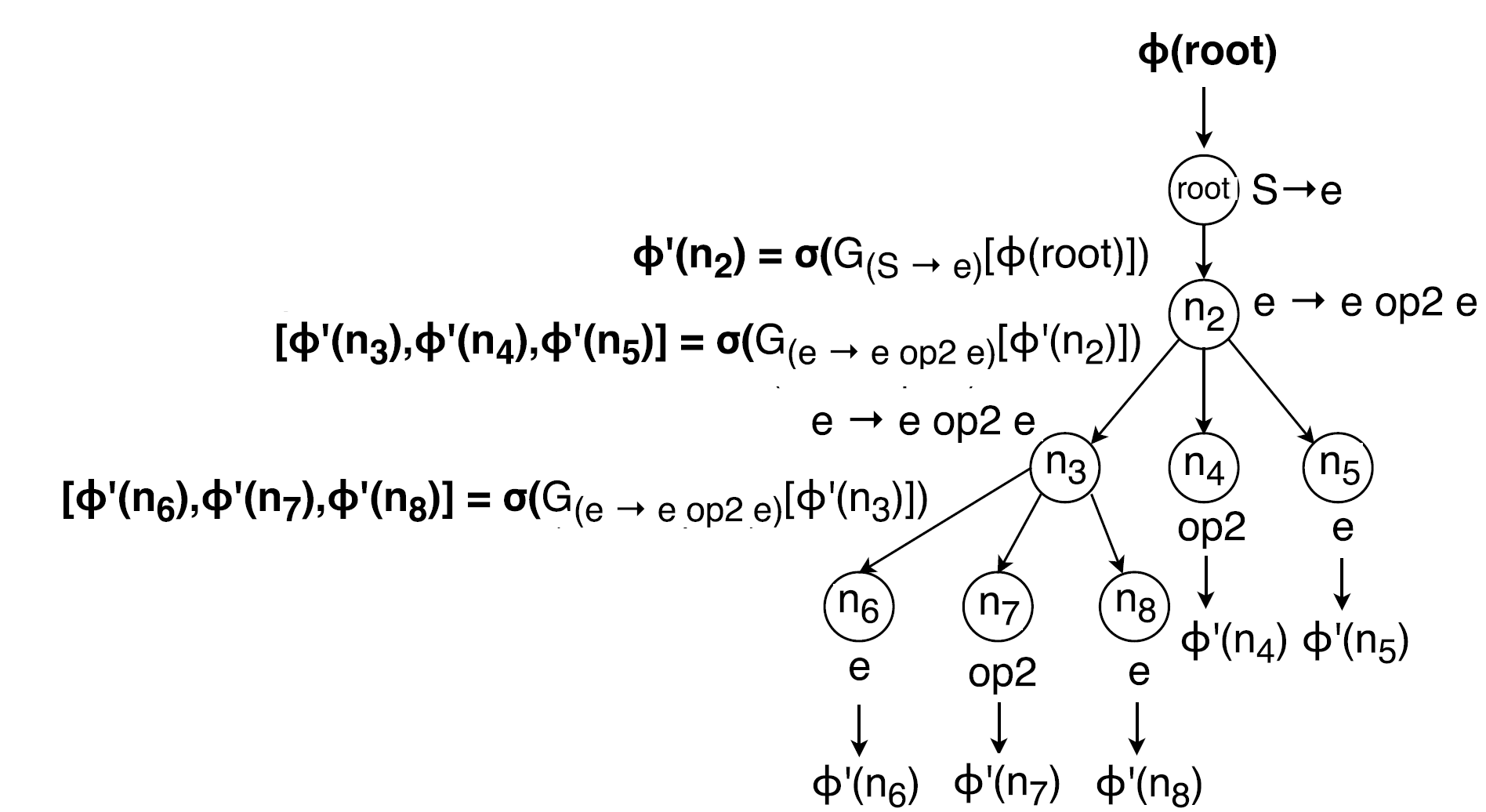}
        \end{minipage}
        \\
        (a) Recursive pass & (b) Reverse-Recursive pass
    \end{tabular}
    \caption{(a) The initial recursive pass of the R3NN. (b) The reverse-recursive pass of the R3NN where the input is the output of the previous recursive pass. Illustrations taken from \citep{nsps}.}
    \label{r3nn}
\end{figure*}

The workings of the R3NN are illustrated in Figure \ref{r3nn}.
The R3NN utilizes two parallel sets of neural networks, $f_r$ and $g_r$,
both consisting of one neural net per grammar production rule $r \in R$.
In the diagram these are denoted by $W(r)$ and $G(r)$, respectively,
with $r$ calculated as production rule $R(n)$ of non-leaf node $n$.
or determined by its symbol $s \in S$ at any leaf $l \in L$,
where $s$ represents an operator in our grammar, calculated by $S(l)$.
These two sets correspond to the R3NN's two passes through an AST (explained below).
Their neural networks use a single layer with hyperbolic tangent activation function (denoted by $\sigma$).

R3NN defines a hyperparameter $M$ indicating the number of features used in their embeddings.
It uses this in an embedded representation $\phi(s) \in \mathbb{R}^M$
for every operator in the DSL (e.g. \verb|+|), which they refer to as \emph{symbols} $s \in S$,
as well as in a representation $\omega(r) \in \mathbb{R}^M$ for every production rule $r \in R$.

The R3NN first makes a \emph{recursive pass} from the embeddings $\phi(l)$
of leaves $l \in L$ of the program tree gradually toward the root node of the partial program,
making for an embedding $\phi(root)$ of the full program so far.
Given a number of child nodes $Q$ of the branch in question
(as dictated by the grammar expansion rule it represents),
this recursive pass goes through neural networks $f_r$ at each branch,
mapping from $Q \cdot M$ to $M$ dimensions,
i.e. from concatenated right-hand side (RHS) vectors to a left-hand side (LHS) vector.

It then performs a \emph{reverse recursive pass} from this root embedding $\phi(root)$,
now passing back to the leaves through $g_r$,
one of a second set of neural networks,
mapping back from $M$ to $Q \cdot M$ dimensions,
i.e. from a LHS vector back to concatenated RHS vectors $\phi'(c)$ for any node $c$,
which now have their individual embeddings instilled with structural information about the entire program and how they fit into this larger structure.

In the event a node $c$ constitutes a non-leaf node $n$,
this process is then repeated, until reaching leaf embeddings $\phi'(l)$.
Such leaf embeddings $\phi'(l)$ are now different for leaf nodes sharing the same symbol,
while before these two passes their original embeddings $\phi(l)$ would have been identical.

\citet{nsps} define an \emph{expansion} $e \in E$ as a combination of a hole (non-terminal leaf node) $e.l$ and a grammar production rule $e.r$, together making up for a way we can expand our PPT.
Expansion scores of an expansion $e$ they define as the dot product of their respective embeddings:
$z_e = \phi'(e.l) \cdot \omega(e.r)$.

These scores are then normalized to probabilities using a \emph{softmax} operation.
They find processing leaf embeddings $\phi'(l)$ by a bidirectional LSTM%
~\citep{huang2015bidirectional} before the score calculation helped as well.
To condition R3NN expansion probabilities on the input-output examples specifying desired program behavior in PBE,
they concatenate them to the node features $\phi$ before the recursive pass.

For the training phase they use the strong supervision (see start of Section \ref{sec:neuralpbe})
setup of supervising by the \emph{task function},
i.e. the function we aim to synthesize,
while for the test phase they sample 100 programs from their trained model.
They consider the synthesized program to have passed if any of these demonstrate the correct behavior on the input/output.

As this is a strongly supervised model,
the loss $J^{(task)}_{PPT}$ predicting one hole-expansion for a task function $task$ given a
partial-program tree $PPT$ is defined as the cross-entropy between our predicted probability matrix
$\mathbf{\hat{P}}_{PPT}$ (over holes $H$ and expansion rules $R$)
versus the golden `probabilities' $\mathbf{P}^{(task)}_{PPT}$ as per the task function we are supervising against,
i.e. a matrix that for each hole is defined as a one-hot vector marking the chosen expansion rule as $1$,
the remaining expansion rules as $0$:

\begin{align*}
J^{(task)}_{PPT} &= \operatorname{CE}(\mathbf{P}^{(task)}_{PPT},\mathbf{\hat{P}}_{PPT}) = -\operatorname{E}_{\mathbf{P}^{(task)}_{PPT}}[\log \mathbf{\hat{P}}_{PPT}] \\
\end{align*}

If the task function consists of more than a single node, then we will obtain such a loss for every such prediction step (each starting from their own $PPT$).
Note that the $PPT$ we start from in a prediction step is carried over to inform our next rule expansion.
Informing our current prediction using previous predictions in such a way makes the model \emph{autoregressive}~\citep{kendall1944autoregressive}.


Some critiques of the model have included it being harder to batch (i.e. enable parallel execution) over multiple task functions for larger programs due to its tree-based architecture,
as well as its pooling at the I/O encoding level being harder to reconcile with
\emph{attention}~\citep{bahdanau2014neural} mechanisms.~\citep{devlin2017robustfill}

For our purposes, by merit of the untyped Microsoft Excel \emph{FlashFill}~\citep{prose} domain this model was tested on,
it also shares a weakness with other neural synthesis models:
as neural synthesis models have usually been applied to untyped domains,
they have not been augmented to use information on types,
while existing type-theoretic synthesis approaches have shown this info to be highly valuable.

\paragraph{Neural-guided search} \label{sec:ngs}

\emph{Neural-guided search} is an another approach to \emph{hybrid} neural synthesis,
and like \citet{nsps}'s \emph{neuro-symbolic program synthesis} model
combines symbolic and statistical synthesis methods~\citep{nps}.
It employs neural components to indicate search order within traditional search methods
such as enumerative search (possibly pruned using deduction),
`sort-and-add' enumeration or sketch filling.~\citep{deepcoder}

\citet{kalyan2018neural} built on this to extend the guidance to each search step,
integrating \emph{deductive} search (e.g. a SAT/SMT solver, extensions
like \emph{conflict-driven} learning~\citep{feng2018program}),
a statistical model judging generalization,
and a heuristic search controller deciding which of the model's suggested branches
to explore (branch-and-bound~\citep{kalyan2018neural}, beam search%
~\citep{polosukhin2018neural}, $A^{*}$~\citep{lee2018accelerating}).
The statistical model mentioned here is where other
neural synthesis methods would fit into this approach.

\citet{feng2018program} expanded on neural-guided deductive techniques
like SMT solvers by \emph{conflict-driven} learning,
ensuring that if e.g. a \verb|map| operation would yield an
output list length not corresponding to the desired length,
other operations suffering from the same issue such as
\verb|reverse| and \verb|sort| would be automatically ruled out as well.

\citet{zhang2018leveraging} focus on incorporating deduced constraints
into the statistical model, to allow taking this info into account
in the decision of which branches to focus on.
Another similar effort has been that of \citet{odena2020learning},
which adds additional features describing properties of a function.
Types however have so far been missing here.

Compared to other neural methods, neural-guided search seems
more of a complementary than a competing effort.
The engineering involved to conciliate the benefits of
different approaches here may be quite involved, and as such
are likely less common in research papers comparing neural components,
but their benefits in production systems seem clear.

\pagebreak

\section{Background} 

In order to test our hypothesis,
we would like to still quickly explain two additional topics before moving on to our own methodology:
\emph{lambda calculus}, which forms the basis for a simple DSL,
as well as \emph{static typing},
which is the topic of our investigation within neural program synthesis.

\subsection{Lambda calculus} \label{sec:lambdacalc}


Whereas modern programming languages might have a broad plethora of grammatical constructs available,
for the purpose of our proof-of-concept we will opt to hide much of this.

Types are most powerful in a setting where the underlying DSL is loosely-constrained,
that is, permits arbitrary type-safe combinations of subexpressions,
such that checks be deferred from the grammar to the type level.
In other words, to keep things simple,
our DSL should ideally support such a notion of an expression,
but preferably as little else as possible.

This brings us to the \emph{lambda calculus}~\citep{lambdacalculus},
the simplest~\citep{selinger2008lecture}
\emph{Turing-complete}~\citep{turing1936computable} grammar
in terms of number of grammar expansion rules.
The lambda calculus requires only three grammatical categories:
variables, function definition, and function application
(in notation further adding parentheses to indicate structure).

With this lambda calculus, we now have a solid basis for a simple expressive grammar that allows us to defer most checks from the grammar to the type level.%
~\footnote{
    As an interesting coincidence,
    using this as the basis of our synthesis target language means
    we will use an implementation of \citet{lambdacalculus}'s lambda calculus
    to address \citet{church1957applications}'s problem of synthesizing programs.
}

\subsection{Complementing synthesis grammars with static typing} \label{sec:statictyping}


One might note here that traditionally,
search spaces in program synthesis have been restricted
primarily using the generally available \emph{context-free grammars},
as in the
\emph{Syntax-Guided Synthesis competition} (SyGuS-Comp)~\citep{sygus},
rather than additionally doing so using types, which may only be available in certain DSLs.
One might wonder: how then might adding restrictions based on types
compare to solely relying on restrictions imposed by a grammar?

In a type system consisting only of \emph{unparametrized types},
such as \verb|boolean| or \verb|integer| but not \verb|list of booleans|,
restraining a search using types is in fact equivalent
to a grammar where types are used as left-hand symbols in the grammar.%

However,
what makes the use of types different from,
and more powerful than grammars in restricting the search space,
is the use of \emph{parametric polymorphism}, i.e. availability of \emph{type variables}:
a function \verb|append| may work using either lists of numbers or lists of strings.
As such, its type signature may be made \emph{generic} such as to have its return type reflect the types of the parameters used.

Having such information available at the type level may add additional information over what is used in the simpler case above.
For example, a function to look up elements in a list based on their respective locations might take as its inputs one list containing any type of element, along with a second list of integers containing the indices.

Now, in a \emph{context-free} grammar,
such distinctions could not be expressed in a meaningful way:
such a grammar would quickly explode to the point of no longer remaining a reasonable abstraction to a human observer.
As such, one may regard the reliance of types over a grammar for the purpose of restricting the search space as a generalization of solely relying on a grammar for the purpose of restricting the search space.%



We may therefore use types to prune out additional programs
that are not sensible, i.e. would not pass a type-check.
This way types may help us restrict the synthesis search space,
as per our hypothesis thereby improving synthesis performance.

\pagebreak

\section{Methodology} 

In this section we will discuss our program synthesis model,
which applies type information to improve synthesis quality in programming by example.

We further explain how this synthesizer builds upon the work of \citet{nsps},
explain the functional programming synthesis DSL we use with this to exploit its features,
as well as how we generate datasets in order to obtain training and test sets in our DSL.

To explain the design decisions we made,
we will go by the synthesizer taxonomy of \citet{gulwani2017program} introduced in Section \ref{sec:litreview}.
The first two criteria, i.e. constraints on expressions of user intent and search space,
together give the background needed to understand both our dataset generation method as well as the synthesizer itself.

We will therefore first explain our design decisions with regard to these,
then continue to lay out the design of our dataset generation method and synthesizer.

One should bear in mind that our goal here is not to create the perfect production-ready synthesizer;
instead we will aim to answer each of these categories with the question:
what is the simplest way in which we might effectively test our hypothesis?

\subsection{User intent}

For our expression of user intent, we would like to use input-output examples,
which may be considered a compromise between what is easier for the end-user,
who may ideally prefer natural-language descriptions of program behavior,
versus what is easier for the synthesizer,
which may ideally prefer a complete formal specification of program behavior.

This puts us in the field of programming by example (PBE),
which has a broad area of application despite being conceptually simple%
.

To (1) reduce ambiguity, (2) increase result quality, and (3) speed up synthesis, a synthesizer may be passed more information in various ways:
\begin{itemize}
    \item additional data within the same mode of user intent, e.g. further input-output examples;
    \item an additional expression of user intent of a different type, e.g. a natural language description~\citep{polosukhin2018neural} or type signature of the desired function~\citep{myth};
    \item more descriptive types~\citep{synquid};
    \item additional features describing properties of the function~\citep{odena2020learning}.
\end{itemize}

Of interest here is the realization that, in modern programming languages, types may be \emph{inferred} even without explicit type annotations.
This is then a hidden benefit of synthesis from input-output examples:
if the types of input-output example pairs may be inferred,
then we may regard this as free additional information we can incorporate in our synthesis process.
    Optionally letting users explicitly clarify their desired function type
    may further help ensure a sufficiently widely-applicable function.

\subsection{Program search space}

The program search space consists of the synthesis language (defined by a \emph{context-free grammar}),
either general-purpose or a \emph{domain-specific language} (DSL),
potentially further restricted to a subset of its original operators,
such as by providing a whitelist of \emph{operators}.

The trade-off here is one of expressiveness (achieve more with less code) versus limiting our search space (ensure we can find a solution within too long).

So, how does this fit into our question on reaching a configuration that could best demonstrate the use of types?
Now, providing empirical evidence on how every design choice impacts the usefulness of type information is not in scope for this thesis,
as just any one good configuration may suffice to demonstrate our hypothesis.
Instead, we will make informed guesses
to pick our language, grammar subset and operator set.

Not having empirical evidence to guide our design decisions upfront, however,
we appear free to take some guidance from program search space considerations to generally improve synthesis efficiency:
how might we achieve the highest amount of expressiveness within a limited search space?
The answers we have found to this question, we argue,
is intuitively in line with a program search space designed to demonstrate the utility in synthesis of type information.

Under this goal, it would seem preferable to pick a limited grammar in the functional programming paradigm.
In the following sections we will lay out how we have reached this conclusion.

\subsubsection{The functional programming context}


The \emph{functional} paradigm, named after the use of functions in mathematics,
has been characterized by its composability or \emph{modularity}~\citep{hughes1989functional},
which is key in the creation of synthesizers that generalize well,
as it encourages reusing existing abstractions to allow for a large expressivity using only a small vocabulary,
matching our synthesizer search space requirement of maintaining expressiveness while limiting our search space.

In addition, functional programming offers various general-purpose programming languages,
which helps potentially make our synthesizer potentially applicable to a wide variety of domains.
It is also well amenable to programming types,
which help reduce the search space in program synthesis.

The basic abstraction in functional programming is the \emph{function}.
This means we would view our synthesized programs as being and consisting of \emph{pure functions}~\citep{fortran95},
i.e. returning a deterministic output for any given inputs,
without performing any additional side effects.%
~\footnote{
    These properties of determinism and lack of side effects are generally taken as prerequisites in \emph{programming by example},
    as we will verify program behavior by comparing the output of our synthesized output to that of our original task function.

    If \emph{non-determinism} came into play,
    forcing us to test samples of our stochastic function output,
    we would need to extend our synthesis domain to something that could well be called \emph{synthesis by property} instead.

    Synthesizing functions with \emph{side effects} instead appears closer to the domain of synthesis from \emph{traces},
    where the synthesis specification instead consists of a description of such triggered side effects as a function of various user inputs.
}


One may well regard programs in this paradigm as constructed of a (nested) tree of function applications.
One reason we would like to consider such programs of a tree-based form,
rather than as a list of imperative statements such as variable definitions or mutations,
is that the view of programs as function compositions guarantees us that any complete type-checking program from the root,
filtered to the right output type,
will yield us output of the desired type,
helping us reduce our synthesis search space to a sensible subset,
devoid of e.g. programs containing variable definitions that end up never being used.

This guarantees that, rather than just branching out,
our search will focus on finding acceptable solutions.
This is to be contrasted with \emph{imperative} programs,
a coding style characterized by variable mutation.
Synthesis for such languages
unfortunately
does not support the use of e.g. type-theoretic approaches,
limiting the synthesis methods we might use there
while giving us less means to constrain our search space.


We will next discuss 
our decision on an actual synthesis language,
followed by a further explanation on how we adapt lambda calculus for our own synthesis grammar.

\subsubsection{Synthesis language}

Our \emph{synthesis} language, or \emph{target} language,
is the language we would like for our synthesizer to generate.
This is as opposed to the \emph{host} language,
i.e. the language that our synthesizer is implemented in.

Neural synthesis methods have often used custom DSLs as the target language,
while for the host language typically using \emph{Python}.
For our purposes working with types however,
it would be nice to be able to defer type logic to an existing language.%

This idea though would require our host language
to be able to construct ASTs for, compile,
and interpret our \emph{target} language,
while also requiring the availability of a
deep learning framework for our model implementations.

We conciliate these requirements
by using \emph{Haskell}~\citep{jones2003haskell} as both our host and target language,
a statically typed, purely functional programming language based on the lambda calculus and featuring type inference,
already in use in various non-neural synthesis papers.%
~\citep{synquid,hornclauses,scythe,gissurarson2018suggesting}%


For a deep learning framework, we evaluated Haskell ports of
\emph{PyTorch}~\citep{pytorch} and \emph{TensorFlow}~\citep{abadi2016tensorflow}.
At the moment of choosing, one Haskell port of PyTorch,
named \emph{HaskTorch}~\citep{hasktorch}%
%
%
,
turned out significantly more active,
and along with its welcoming and helpful community solidified our choice.

\subsubsection{Grammatical subset} \label{sec:grammar}

Our synthesis DSL
only requires a subset of the functionality in the lambda calculus,
that is, function application and referencing variables,
though without function definition.

For the purpose of expressing \emph{partial} programs,
one feature missing in the lambda calculus that we will need to add in our DSL is that of \emph{holes},
i.e. placeholder nodes in the AST to be filled by the synthesizer.


An attempt to express our DSL
as a \emph{context-free grammar},
taking inspiration from the notation of \emph{extended Backus–Naur form} (EBNF)~\citep{standard1996ebnf},
might look as follows:
~\footnote{
    One may note that function application here is \emph{unary} in its number of parameters, as it is in Haskell,
    meaning that multiple-parameter functions must be emulated using a chain of such applications.
    \citet{nsps}'s R3NN, however, presumes nodes may in fact have multiple child nodes.

    Using the R3NN on our DSL then means that we must break \citet{nsps}'s original
    assumption that each branch node itself corresponds to one rule expansion,
    as rule expansions may in our case then span multiple AST branch nodes.
    Despite this shift, the theory of R3NN still applies, however.
}

\begin{verbatim}
expr = "(", expr, ") (", expr, ")";
expr = <any variable contained in our whitelisted operators>;
\end{verbatim}

As such, given an operator list consisting of operators \verb|and| and \verb|false|,
we would then obtain the following EBNF:

\begin{verbatim}
expr = "(", expr, ") (", expr, ")";
expr = "and" | "false";
\end{verbatim}


Now, in practice, we would like to support the use of different operator sets
rather than just the one hard-coded for illustrative purposes above,
so it is fortunate we did not need to fix these at the grammar level itself.

However, this simple grammar can unfortunately still generate some bad programs:
\begin{itemize}
    \item programs where the argument of a function is not of the right type.
        This class of mistakes we are no longer able to reliably solve
        at the grammar level due to our polymorphic parametrism.
        Instead, we wish to defer this kind of check to the type level.
    \item programs where the arity of a given operator is not respected, i.e.
        by invoking more function applications than we up-front know are supported.
\end{itemize}

The latter problem we can deal with in either of two ways:
\begin{itemize}
    \item we ignore the problem by deferring it to the type-level,
    providing a solution consistent with how we handle problems of the former type.
    \item we reframe the grammar by statically \emph{unrolling} any provided operators such as to ensure only valid function arities are supported.
\end{itemize}

We consider the second option to be preferable from the perspective of a type-based synthesizer;
although this would expand the number of production rules in the grammar,
different numbers of function application for a given operator yield different \emph{result types}.
For a type-based synthesizer, distinguishing these makes sense,
as this distinction should facilitate learning.%



    While this poses limitations in terms of supporting arity-agnostic operators
    such as the \emph{argument flipping} and \emph{function composition} combinators,
    we will consider this as sufficient for the purpose of our present paper.


As an example, let's say our operator list again contains the two operators from above,
one operator false, a variable that does not describe a function, and thus takes no arguments,
and one operator \verb|and| as a function as per lambda calculus \emph{curried} to take at most two arguments,

Such a \emph{curried} function allows arguments to be applied one at a time.
The way this works is that, when an argument is applied to a curried form of a function taking two parameters,
the result is a function that still takes one parameter, before yielding the actual result of the original function.

Our \emph{unrolled} grammar would then look as follows:%
    


\begin{verbatim}
expr = "(and ", expr, " ", expr, ")";
expr = "(and ", expr, ")";
expr = "and";
expr = "false";
\end{verbatim}

It must be noted that context-free grammars like the above
describe how to generate \emph{full programs} in the associated grammar.
However,
in \emph{partial programs},
we express \emph{holes} or unresolved \verb|expr| symbols as
using the \emph{dummy variable} of \verb|undefined|
(which in our Haskell context passes compilation unlike the built-in hole `\_'):

\begin{verbatim}
    expr = "(and ", expr, " ", expr, ")";
    expr = "(and ", expr, ")";
    expr = "and";
    expr = "false";
    expr = "undefined :: ", <type>;
\end{verbatim}

This grammar is not technically a valid EBNF,
as we have deferred specifying the type.
This is not a coincidence: the type actually depends on the entire program tree.
In other words, our grammar productions are dependent on context,
meaning our grammar is not actually \emph{context-free}.

As such, context-free grammar notations such as EBNF cannot
fully express our production rules inclusive of hole types.
Any implementation however might use the synthesis language's type inference,
in our case built into Haskell, in order to calculate these types.




\subsubsection{Operator whitelist}

Our synthesis approach itself is agnostic to the set of operators used,
allowing for relatively straight-forward experimentation with different sets of operators.
Adding new operators simply involves generating a new dataset, then retraining the model.


We will further expand on our operator set
in Section \ref{sec:experiment}.

\subsection{Dataset generation} \label{sec:datagen}

As we were unable to find existing datasets in the functional program synthesis domain
of a size appropriate for training a neural model,
we have opted to instead generate a dataset of our own.

As the potential space of viable programs is potentially unbounded,
we instead opt to artifically limit the space to generate from.

Our main goal in creating a dataset consists of generating the programs to be synthesized,
alongside the input-output data we would like to use to synthesize them from (as per our PBE setting).
Now, the inputs here are generated, whereas the outputs are obtained simply by running these inputs through our programs.

However, as our programs may take parameters of parametric types, e.g. list of any given type \verb|[a]|,
we take the intermediate step of instantiating such types to monomorphic types,
i.e. types not containing type variables themselves,
which we may then generate inputs for.

Note that to make our task easier,
we further maintain such a separation by type instances for our generated programs,
meaning that a potential \emph{identity function} in our dataset might be
included in our training set under type instance $Int \rightarrow Int$,
then perhaps in our test set under another type instance like $Char \rightarrow Char$.
%
We may sometimes still refer to just task functions however, as the distinction is not otherwise relevant.

An example showing what different components of our dataset items might look like may be found in Figure \ref{fig:datasample}.

\begin{figure*}
    \begin{tabular}{|l|l|} \hline
        \textbf{task function} & \verb|let just = Just; compose = (.) in compose just unzip| \\ \hline
        \textbf{type instance parameter input types} & \verb|[(Int, Char)]| \\ \hline
        \textbf{type instance output type} & \verb|Maybe ([Int], [Char])| \\ \hline
        \textbf{input expression} & \verb|([((17), '0'), ((20), '2')])| \\ \hline
        \textbf{output expression} & \verb|Right (Just ([17, 20], "02"))| \\ \hline
    \end{tabular}
    \caption{A task function instance from our dataset with a corresponding sample input/output pair.}
    \label{fig:datasample}
\end{figure*}

Our full generated dataset consists of the following elements:
\begin{itemize}
    \item the right-hand symbols or operators we allow in our DSL, to be detailed in Section \ref{sec:task};
    \item the types of any task function in our dataset;
    \item sample input-output pairs for different type instances of our task functions;
    \item a split over training/validation/test sets of any of our tasks, i.e. type instances for a given task function;
    \item pairs of symbols in our DSL with their corresponding expansion rules (including type annotations for holes);
    \item types of any expansion rules in our DSL; 
    \item NSPS's maximum string length $T$, based on our stringified input-output examples (also taking into account types for the augmented model);
    \item mappings of characters to contiguous integers so we can construct one-hot encodings covering the minimum required range of characters (tracked separately for input-output, types, and either);
    \item the configuration used for data generation to make data reproducible, discussed further in Appendix section \ref{sec:gen-params};
    \item the types we generate to instantiate type variables, again for reproducibility purposes, separated by arity based on the number of type parameters they take.
\end{itemize}

A brief overview of how to generate such a dataset to train our synthesizer on is shown in Algorithm \ref{alg:gen}.

\begin{algorithm}
    \caption{dataset generation}
    \label{alg:gen}
    \begin{algorithmic}
        \State \textbf{given}: expression space $E$, operators or symbols $s \in S \subset E$, expansion rules $r_s \in R \subset E$, programs $p \in E$, types $t \in T$, monomorphic types $t^{(m)} \in T^{(m)} \subset T$, input expressions $i \in E$, output expressions $o \in E$, parameters $a$;
        \State \textbf{calculate} expansion rules $r_s^{(1, \dots, n)}$ \textbf{from} $s \in S$ \textbf{by} unrolling our grammar symbols; 
        \State \textbf{generate} any possible program $p$ \textbf{given} expansion rules $\forall s : r_s^{(1, \dots, n)} \in R^n$ and a max number of holes;
        \State \textbf{sample} monomorphic types $t^{(m)} \in T^{(m)}$ \textbf{up to} a max number and \textbf{within} a given nesting limit;
        \State \textbf{generate} instances $t^{(m)}_{a_p^{(1, \dots, n)}}$ \textbf{for each} generic non-function parameter types $\forall p : t_{a_p^{(1, \dots, n)}}$ \textbf{given} sampled types $t^{(m)}$;
        \State \textbf{sample} type instances $t^{(m)}_p$ \textbf{for each} function type $\forall p \in E: t_p$ \textbf{up to} a given number;
        \State \textbf{generate} sample expressions $i_{t^{(m)}_{a_p^{(1, \dots, n)}}}^{(1, \dots, n)}$ \textbf{for each} non-function parameter type instance $t^{(m)}_{a_p^{(1, \dots, n)}}$, \textbf{up to} a maximum each and \textbf{within} given value bounds;
        \State \textbf{calculate} a filtered map of generated programs $p^{(1, \dots, n)} \in E$ \textbf{for each} instantiated function parameter type combination $\forall a_p : t^{(m)}_{a_p^{(1, \dots, n)}}$ \textbf{by} matching its type to obtain samples $i_{t^{(m)}_{a_p^{(1, \dots, n)}}}^{(1, \dots, n)}$ for our function types;
        \State \textbf{calculate} outputs $o_{t^{(m)}_p}^{(1, \dots, n)}$ \textbf{for each} task function instance $t^{(m)}_p$ \textbf{given} a sample of generated inputs $i_{t^{(m)}}^{(1, \dots, n)}$;
        \State \textbf{filter} out program type instances $t^{(m)}_p$ \textbf{without} i/o samples $(i,o)_{t^{(m)}_p}^{(1, \dots, n)}$;
        \State \textbf{filter} out any functions instances $t^{(m)}_p$ \textbf{with} i/o behavior identical to others to prevent data leakage;
        \State \textbf{sample} task function type instances $t^{(m)}_p$ \textbf{from} any remaining programs $p$;
        \State \textbf{calculate} longest strings and character maps;
        \State \textbf{split} our task function type instances $t^{(m)}_p$ \textbf{over} train, validation and test datasets.
    \end{algorithmic} 
\end{algorithm}


We first generate our expansion rules by \emph{unrolling} each operator in the dataset
as described in Section \ref{sec:grammar},
using a different number of holes corresponding to any applicable arity.


To create our dataset of task functions,
we start from an expression consisting of only a hole,
then step by step generate any type-checking permutation
by filling a hole in such an expression using our expansion rules.
We only fill holes in a generated expression up to a 
user-defined limit,
disregarding any programs still containing holes after this point.

Like \citet{nsps} we uniformly sample programs from our DSL,
based on a 
user-defined maximum,
while still respecting the above complexity limits.
We similarly use sampling for the generation of sample input-output pairs and,
for instantiating our type variables, monomorphic types,
i.e. types not containing type variables.

While we quickly mentioned type-checking programs to filter out bad ones,
we had yet to expand on this practice:
we presently use a Haskell interpreter to type-check our generated programs at run-time,
filter out non-function programs (e.g. \verb|false|),
and check if program types look \emph{sane}:
to weed out some programs we deem less commonly useful,
we filter out types \emph{containing} functions (e.g. list of functions),
as well as types with constraints that span more than a single type variable (e.g. $(Eq (a \rightarrow Bool)) \Rightarrow a$).%
~\footnote{
    Programs not passing these checks are not necessarily invalid,
    but by our engineering judgement,
    are much more circumstantial in their usage,
    making for only a smaller portion of valid programs,
    aggravating our search space problem.
    For this reason, we would currently prefer for our synthesizer to focus on
    the region of our search space that we generally deem to be of higher interest.
}

As we cannot directly generate samples for types containing type variables,
we first instantiate any such type variables using a fixed number of monomorphic types we generate.
We define a
maximum level of type nesting for such sampled types,
to prevent generating types like `list of lists of booleans'.
We further specify a maximum number of types generated.

We then use these monomorphic types to instantiate any polymorphic
(non-function) input types occurring in our task functions.
To simplify things, we restrict ourselves to substituting only non-parametric types (e.g. boolean yet not list of boolean) for type variables contained in a larger type expression.
In the event the type variables in our types involve \emph{type constraints},
we ensure to only instantiate such type variables using our monomorphic types that satisfy the applicable type constraints.

This yields us a set of monomorphic input types,
for which we then generate up to a given maximum number of sample inputs,
although this may get less after filtering out duplicate samples.
We use hyperparameters to indicate range restrictions for different types here.

For any given given task function type signature,
we then check for the types of each of their input parameters,
and take any corresponding combination of type instances in case of polymorphic types.

Now, for any non-function parameter types,
we may just take the previously generated sample input-output pairs for those types.
Parameters with function types, however,
we instead instantiate to function values by just taking
any of our generated task functions corresponding to that type.

Based on these sample inputs, we would then like to generate corresponding outputs for our generated task functions.
For our task functions that are polymorphic, i.e. contain type variables,
we must do this separately for different type instances.


We run our programs using our run-time Haskell interpreter.
We catch run-time errors on specific inputs such that
we can regard these errors as just another resulting output
that our synthesizer should consider when comparing behavior between programs.
In other words, a \emph{partial function},
i.e. a function that only works on a subset all inputs of the desired input types,
may still constitute a valid program that we may wish to learn to synthesize.

Having generated input/output examples for our task functions,
we finally filter out any task function type instances for which we have somehow failed to generate such samples.
We moreover limit our dataset to a given maximum.

At this point we:
\begin{itemize}
    \item use a random split to divide our task function type instances over training, validation and test sets;
    \item calculate the longest input-output examples in our dataset (as string), when considering types (as per our experiment) also taking into account the length of the string representations of such types of inputs and outputs;
    \item track any characters used in string representations of the expressions in our dataset
    (for our type experiment also those used in string representations of the types),
    and assign them to indices for our \emph{one-hot} encodings of input-output examples (and their associated types).
\end{itemize}

\subsubsection{Preventing data leakage}

One additional concern here is \emph{data leakage}~\citep{leakage},
which is the issue of a model being able to `cheat' on its predictions due to e.g. items shared between training and test sets.

Now, in \emph{supervised learning} settings,
one would typically ensure labeled samples would not be duplicated across sets:
if we train a classifier on a cat picture,
then evaluate it on this same cat picture,
the classifier could simply remember the examples,
rather than learning how to generalize in its task to \emph{unseen} samples.

In \emph{programming by example} on the other hand,
the equivalent is not simply to ensure \emph{task function} instances are deduplicated across sets.
Imagine we had two distinct task function instances exhibiting identical behavior:
if one were allocated to our training set, the other to the test set,
then we have again created an opportunity for our synthesizer to cheat:
it could remember the program from our training set,
then synthesize it during evaluation.

While this program would not be the same,
for the purpose of synthesis \emph{evaluation} of accuracy,
synthesizing a function exhibiting the correct behavior already qualifies as successful synthesis,
as synthesizing one specific function implementation is not the goal of programming by example.

    Unfortunately, under the \emph{strong supervision} used in our implementation,
    these two are conflated for the purpose of calculating the loss.
We avert this problem by ensuring no task function instances across different datasets share the same \emph{input-output pairs}.

Now, one would then presume that as long as we then ensure
that functions across different datasets
would not share the same \emph{input-output pairs},
this issue would be averted.

However, the reality is slightly more complicated still:
presume we have an \verb|increment| function mapping input $0$ to $1$,
along with a \verb|successor| function mapping $0$ to $1$ and \verb|False| to \verb|True|.

If we would simply deduplicate by identical input-output pairs,
we would conclude these functions to behave differently,
and would accept e.g. having the \verb|successor| function in our training set,
the \verb|increment| function in our test set.

This again leads us to a similar same problem however:
our synthesizer could simply memorize how to synthesize our \verb|successor| function,
then during evaluation use this knowledge to pass our \verb|increment| synthesis test.
As such, the solution would be to ensure we cannot train on task functions exhibiting all of the behaviors of a function we would evaluate on.

The way we tackle this in our implementation is to identify any such function pairs where either would fully subsume the behavior of the other.
For the sake of simplicity, we presently ensure that only one of any such semi-duplicate set is kept in our dataset,
rather than still allowing less general versions in our training set.

When deciding which task function instance of a similar pair to keep,
we first look for the more general function (i.e. operating across more type intances as used in our dataset),
otherwise look for the task function with the shortest implementation (in terms of number of nodes),
or finally, as a tiebreaker, arbitrarily keep either of the two.

\subsection{Search technique}

\subsubsection{Our adaptation of neuro-symbolic program synthesis} \label{sec:ournsps}

Type-based approaches typically call for a \emph{top-down search strategy}.
As such, we will need to build upon tree-based (or AST-based) rather than sequential (or token-based) neural synthesis methods:
we can apply types to an AST containing holes,
which we cannot do for arbitrary sequences representing a partial program.

As a benchmark algorithm we will therefore use the neuro-symbolic program synthesis
method from \citet{nsps} introduced in Section \ref{sec:nsps},
a top-down neural synthesis method.








For conditioning programs we use a (bidirectional) LSTM.
We also use \citet{nsps}'s bidirectional LSTM processing global leaf representations right before score calculation.

Like them we also use the \emph{hyperbolic tangent} activation function in the neural networks part of the recursive and reverse recursive passes of the \emph{R3NN}.
As a place for input-output conditioning we use \emph{pre-conditioning}
(adding embedded input-output examples before the recursive pass),
which they report to work best.

While not indicated in their paper but only in a related patent~\citep{mohamed2017neural},
it appears the synthesizer was trained using the \emph{Adam} optimizer~\citep{kingma2014adam}.
We follow this example.

As in the original paper,
our evaluation on the test set involves sampling 100 programs using the synthesizer (which may include duplicates),
and considering the trial a success if any of these demonstrates the desired program behavior.
A more sophisticated alternative here might be to use a controller such as \emph{beam search}~\citep{polosukhin2018neural}.


For the purpose of calculating total loss for an epoch,
we presently aggregate the loss over the different
prediction steps by taking the mean of their respective losses,
then similarly aggregate over task function instances in our training set.
Such losses across prediction steps for a task function instance
potentially cover a hole multiple times if it initially remains unfilled.

\subsubsection{Functional program domain} \label{sec:fp}

Aside from the grammar we have described in Section \ref{sec:grammar},
translating \citet{nsps}'s synthesizer from its original \emph{FlashFill}~\citep{prose} domain to our domain of \emph{functional programs},
we have had to make the following adjustments to their original algorithm:

    While the input-output samples used by \citet{nsps} were all \emph{strings},
    in our functional domain these could essentially comprise arbitrary expressions.
    While ideally a synthesizer would respect the tree-like structure of such expressions as ASTs,
    our naive approach has been to simply perform \emph{sample serialization} here,
    taking string versions of our actual input/output expressions,
    then \emph{one-hot encode} the strings' characters as \citet{nsps} did using their string samples.

    In our functional setup,
    we distinguished potentially multiple \emph{type instances} for each function,
    which may depend on its type signature's number of \emph{type variables},
    as well as on any potential \emph{type constraints} on these type variables.


    While this does not pose a problem for our sample encoder,
    it did actually end up problematic for the purpose of \citet{nsps}'s \emph{R3NN},
    which conditioned programs on these sample features using an LSTM,
    which expected a fixed number of embedded samples.
    To work around this,
    we sample a fixed number of i/o pairs per task function instance during dataset generation.

    We have chosen to sample \emph{without replacement} for any
    sample sizes lower than the number of available input-output example pairs,
    which offers a lower stochasticity than sampling with replacement.
    In the event we have less pairs available (e.g. 4) than we would like to sample (e.g. 10),
    we would first take each available pair while we still wish to sample more items than are available in the pool (i.e. first sample the 4 available pairs twice),
    then finally use sampling without replacement to uniformly sample the remaining desired pairs (i.e. randomly pick 2 of the 4).

\subsubsection{Types} \label{sec:typednsps}

We will now explain how we augment the NSPS model to incorporate type info.
We also refer back to NSPS's hyperparameters $T$, $H$ and $M$ here,
where $T$ indicates the maximum string length for any input or output string,
$H$ controls the amount of features per one-hot encoded characters,
$M$ indicates the number of features the \emph{R3NN} uses in its embeddings,
$\omega(r)$ the production rule embeddings,
and $\phi(s)$ the symbol embeddings.

Consistent with how we embed expressions, we similarly serialize types to strings,
then one-hot encode their characters as we do for input/output expressions.
To get the most out of our types, we will want to provide them for:
\begin{itemize}
    \item inputs and outputs,
    which we simply incorporate as additional features in \citet{nsps}'s
    \emph{example encoder} as explained in Section \ref{sec:nsps},
    concatenating their one-hot embeddings to those of the input/output pairs before passing them through the input/output LSTMs,
    increasing the amount of features per sample under their baseline
    LSTM encoder by another $4HT$ making for a total of $8HT$ features per sample;

    \item expressions from expansion rules $r$; 
    for these we may calculate types statically upfront%
    %
    %
,
    then embed these to obtain $M \cdot T$ features per expansion rule $r \in R$,
    and during R3NN prediction concatenate these features to the existing
    representation $\omega(r) \in \mathbb{R}^M$,
    yielding $\omega'(r) \in \mathbb{R}^{M \cdot (T+1)}$.

    \item (hole) AST nodes $c$ in any PPT.
    %
    %
    A proper attempt here would be based on type inference across the program.

    For the sake of simplicity, however, we will settle for simply using the hole's
    parent branch node to obtain its parameter type without type variables filled out,
    i.e. a \emph{local} type that has yet to take into account some of the type information available elsewhere in the PPT.

    During prediction in the R3NN, we then embed these types by an LSTM into $M \cdot T$ features per hole type.
    As with rule embeddings, we then concatenate these with the original $M$ hole node features,
    once concatenated together, yielding $\phi''(l) \in \mathbb{R}^{M \cdot (T+1)}$.

\end{itemize}

Having obtained our respective rule and hole embeddings expanded to $M \cdot (T+1)$ from the original $M$ features,
we would then proceed to calculate the scores from these enhanced embeddings using the same calculations as before,
simply swapping out the embeddings to their enhanced versions,
i.e. going from $z_e = \phi'(e.l) \cdot \omega(e.r)$ to $z_e = \phi''(e.l) \cdot \omega'(e.r)$.

The basic idea here is simple:
on the type level we may provide information not only about how outputs correlate to inputs,
but may also provide info about how our expansion rules may match up to particular holes in our program.
Using this extra information should improve our search,
as per our research hypothesis. 

\pagebreak

\section{Experiment} \label{sec:experiment}

We will now detail the particular setup we used for our experiment.
We will first explain the benchmark task we use in our experiment in
the next section.
To perform our experiment, we
first find an appropriate learning rate on our vanilla implementation of NSPS,
otherwise taking the hyperparameter values described in Section \ref{sec:hpar}.

We will then evaluate on our task to evaluate a few different models;
our vanilla implementation of \citet{nsps}'s NSPS model,
our type-based additions described in \ref{sec:typednsps},
as well as an enlarged version of the vanilla model for fair comparison.
To reduce variance, we run each model to convergence using $4$ different seeds.
For final evaluation, we provide a uniform random synthesizer as a reference baseline as well.

\subsection{Benchmark task} \label{sec:task}

We pick our own set of types and operators to generate a dataset as described in Section \ref{sec:datagen}.
For this purpose we have picked a limited set of operators widely applicable over the types used.
To this end, we focus on functions operating on \emph{typeclasses} based on categories taken from \emph{category theory}.
Types used and their membership to our used typeclasses may be found in Figure \ref{typeclasses}.

\begin{figure*}
    \begin{tabular}{|c|c|c|c|c|c|c|} \hline
        $\downarrow$ typeclass / dataclass $\rightarrow$ & Char & Int & Maybe & List & (,) & Either \\ \hline
        Enum & \textbigcircle & \textbigcircle & & & & \\ \hline
        Foldable & & & \textbigcircle & \textbigcircle & \textbigcircle & \textbigcircle \\ \hline
        Traversable & & & \textbigcircle & \textbigcircle & \textbigcircle & \textbigcircle \\ \hline
        Functor & & & \textbigcircle & \textbigcircle & \textbigcircle & \textbigcircle \\ \hline
        Monoid & & & \textbigcircle & \textbigcircle & & \\ \hline
        Semigroup & & & \textbigcircle & \textbigcircle & \textbigcircle & \textbigcircle \\ \hline
    \end{tabular}
    \caption{The types used in generating our synthesis dataset (column headers), along with their respective \emph{typeclass} memberships as marked by a `\textbigcircle'.}
    \label{typeclasses}
\end{figure*}

In order to pick our operator set, 
we had to use a number of heuristics to filter down the operators (comprising either functions or \emph{data constructors}) offered:
\begin{itemize}
    \item avoiding \emph{side-effects},
    as this category is not a great match for \emph{programming by example};
    \item avoiding unnecessarily \emph{type-specific} operators, as we would like to keep our operator set small yet reusable;
    \item focusing on different \emph{inter-type} over using various similar \emph{intra-type} operators, as the former helps make our operator set more amenable to reducing the search space using type information;
    \item avoiding \emph{infinite lists}, which are relatively circumstantial in their usage, as these would crash evaluation unless specifically converted back to finite lists;
    \item functions particularly circumstantial in their accepted input values,
    e.g. the list-indexing \verb|take| function requiring input integers that are valid index values of the given list;
    \item functions taking \emph{predicates}, i.e. functions returning \emph{boolean} values, as we have few predicate functions we can use to this end given the previous point, as many of these require concrete values as parameters;
    \item function variants using a parameter order less amenable to \emph{currying};
    \item functions \emph{discarding values}, as these mostly make for redundant programs;
    \item
    functions otherwise likely to induce redundancy, such as the \emph{identity function};
    \item reducing \emph{redundancy}, i.e. preferably not including operators that can already be constructed from existing ones;
    \item preferably reducing the number of functions with \emph{similar type signatures};
    \item not necessarily aiming for full \emph{expressiveness} in terms of manipulation of the chosen types.
\end{itemize}

Inspired by this list of heuristics,
we have picked the following set of operators for our chosen types:
\verb|0|, 
\verb|zero|, 
\verb|Just|, 
\verb|maybe|, 
\verb|(:)|, 
\verb|length|, 
\verb|(,)|, 
\verb|zip|, 
\verb|unzip|, 
\verb|toEnum|, 
\verb|fromEnum|, 
\verb|foldMap|, 
\verb|elem|, 
\verb|sequenceA|, 
\verb|sequence|, 
\verb|fmap|, 
\verb|mempty|, 
\verb|(<>)|, 
and \verb|(.)|%
. 

\pagebreak

\section{Result} \label{sec:result}

Having added our type-level supervision during training, we expect synthesis success rates to rise
compared to the baseline algorithm.
This demonstrates that the findings from traditional program synthesis methods are relevant also in the field of neural program synthesis.

Any results here are trained on our dataset spanning programs of up to 3 nodes,
during training evaluated by sampling 100 programs from the synthesizer for any task function instance.

\begin{figure*}[h]
    \includegraphics[scale=0.7]{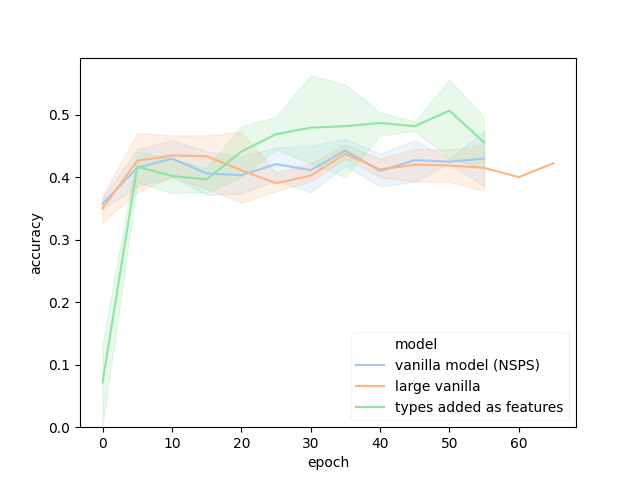}
    \caption{
        Prediction accuracy over 100 samples across training epochs for our different model variants,
        trained on our dataset of programs of up to 3 nodes.
    }
    \label{fig:accuracy}
\end{figure*}

Accuracy results over training on our validation set can be found in Figure \ref{fig:accuracy},
while accuracy on the test set for the fully trained models may be found in Figure \ref{fig:finalacc}.
We see that the simpler vanilla NSPS model is both more limited in variance while also learning to a more limited extent before converging.

Our first observation here is that the task has turned out relatively challenging,
with accuracy for the baseline model increasing only somewhat beyond its initial random accuracy.
Furthermore, most of the gains in accuracy for the baseline model are attained over the initial 10 epochs of training.
We feel these issues may be explained largely from the limited size of our dataset.

We additionally trained an enlarged version of this baseline model,
doubling the sample encoder output dimension parameter $H$ from $32$ to $64$,
giving it a similar amount of expressiveness as our typed model as a reference.
To our surprise however, this model
fared little better than the baseline, again likely stemming
from generalization issues related to the size of our dataset.

Our `typed' NSPS model however,
keeping $H$ at $32$ but allotting that same amount for types,
starts from sub-random accuracies, yet ends up able to learn more,
after $20$ epochs out-performing both our baseline and enlarged models,
indicating it is in fact worthwhile to distribute features between input/output pairs and types.

\begin{figure*}
    \begin{tabular}{|l|c|c|c|c|c|c|c|c|c|c|} \hline
        & \multicolumn{5}{|c|}{ \textbf{evaluated @ 20 samples} } & \multicolumn{5}{|c|}{ \textbf{evaluated @ 100 samples} } \\ \hline
        & \multicolumn{2}{|c|}{ \textbf{accuracy} } & \multicolumn{3}{|c|}{ \textbf{acc mean @ x nodes} } & \multicolumn{2}{|c|}{ \textbf{accuracy} } & \multicolumn{3}{|c|}{ \textbf{acc mean @ x nodes} } \\ \hline
        \textbf{experiment} & \textbf{mean} & \textbf{var} & \textbf{1} & \textbf{2} & \textbf{3} & \textbf{mean} & \textbf{var} & \textbf{1} & \textbf{2} & \textbf{3} \\ \hline
        \textbf{vanilla NSPS} & 0.13 & 0.000 & 0.27 & 0.13 & 0.11 & 0.37 & 0.000 & 0.77 & 0.36 & 0.30 \\ \hline
        \textbf{large} & 0.12 & 0.001 & 0.30 & 0.12 & 0.09 & 0.38 & 0.002 & 0.73 & 0.38 & 0.30 \\ \hline
        \textbf{typed} & \textbf{0.22} & 0.002 & \textbf{0.55} & \textbf{0.24} & \textbf{0.12} & \textbf{0.48} & 0.003 & 0.77 & \textbf{0.55} & \textbf{0.34} \\ \hline
        \textbf{uniform random} & 0.14 & 0.000 & 0.43 & 0.11 & 0.11 & 0.33 & 0.000 & \textbf{0.84} & 0.25 & 0.31 \\ \hline
    \end{tabular}
\caption{
    Summary of final prediction accuracy on our test set over different models after training (4 seeds each),
    for each selecting the best-performing epoch as per validation accuracy i.e. before overfit.
    Results are separated by evaluation on 20 vs. 100 samples,
    and metrics include mean accuracy, variance of accuracy,
    as well as mean accuracy separated by programs of a given number of nodes. Bolded cells indicate the best accuracy in a given category.
}
\label{fig:finalacc}
\end{figure*}

We additionally attempted to look into whether we could find a correlation between this
advantage gained by incorporating type information (as seen in Figure \ref{fig:ttest}) with the number of nodes in a program.
However, the data we managed to gather was insufficient to demonstrate such a correlation.


Specifically, our task turned out hard,
with none of the models significantly out-performing a random synthesizer
on the task functions of 3 nodes.
On the 1-node task functions at 100 samples,
similarly none of the models outperformed the random synthesizer's 84\% accuracy,
although at 20 samples the typed model out-performed it at 55\% vs. 43\%.
This means that most of the learning success for these models has been on the 2-node task functions,
particularly for the typed model, achieving over double the random accuracy for either 20 or 100 samples.

Unfortunately for our purposes, performing best on the 2-node programs
means we cannot presently distinguish any clear correlation between
task function node size and advantage from type information.
As such, we would love to see similar experiments scaled up to larger datasets including higher node sizes.

In order to verify the statistical significance of our proposed model variant,
we have performed an independent two-sample t-test comparing the accuracy across our models
(as evaluated on 20 samples per task function instance),
results for which may be found in Figure \ref{fig:ttest}.

Whereas we similarly compared such differences for accuracy as evaluated on 100 samples,
we found the outcome for 20 samples to more clearly demonstrate the difference between models.
Under 4 seeds each our typed model turns out to have only p=1.0\%
to stem from the same distribution as our baseline model,
indicating the improvement in performance is statistically significant.

\begin{figure*}
    \begin{tabular}{|l|c|c|c|c|c|c|} \hline
        \textbf{p-values}
        & \textbf{vanilla} & \textbf{large} & \textbf{types} & \textbf{uniform} \\ \hline
        \textbf{vanilla} & 1.000 & 0.859 & 0.010 & 0.013 \\ \hline
        \textbf{large} & 0.859 & 1.000 & 0.024 & 0.065 \\ \hline
        \textbf{types} & 0.010 & 0.024 & 1.000 & 0.001 \\ \hline
        \textbf{uniform} & 0.013 & 0.065 & 0.001 & 1.000 \\ \hline
    \end{tabular}
    \caption{P-values for different model combinations under the independent two-sample t-test comparing accuracy as evaluated on 100 samples per task function instance.}
    \label{fig:ttest}
\end{figure*}

\pagebreak

\section{Discussion} 

\subsection{Dataset considerations}

In Section \ref{sec:result} we mentioned dataset design as one of our concerns,
as the size of our current dataset potentially may have hindered further generalization,
consisting of only 259 task function instances in total,
split up across training, validation and test datasets.

Picking our dataset definitely felt like a dilemma.
The dataset we used supported task functions of up to 3 nodes.
We initially tried sampling only 20 programs upon evaluation to reduce running times,
but to generate reliable accuracy graphs raised this to sampling 100 programs,
which meant that 95\% of training time was in fact spent on accuracy evaluations.

This put us in the undesirable position of having to choose
between dataset sizes not conducive to strong generalization,
versus increasing run-time and memory requirements,
as an increase in the maximum number of nodes brings a potentially exponential growth in the size of the dataset.
Facing a compiler-induced memory leak through our interpreter library~\citep{hintleak},
we felt compelled to err toward our current dataset of
limited size to keep run-time requirements in check.

We had in fact constructed a dataset of programs of up to 4 nodes,
for which we were in fact able to confirm a lower level of overfit than on our current dataset,
but had not evaluated on 100 samples during these runs,
preventing us from generating reliable accuracy graphs for this.

However, another potential culprit to proper generalization
would be our \emph{strong supervision} setup itself,
which would consistently penalize any correct solutions
not identical to the task function itself.
While in a larger dataset these biases may even out,
at our scale this might well still pose a barrier to learning.

\subsection{Design limitations}

\begin{itemize}
    \item
    Our implementation settled for \citet{nsps}'s \emph{baseline LSTM encoder} rather than its more complex \emph{cross-correlation} encoder variants.
    While such variants may improve synthesis quality,
    these come at an increased amount of computation as well.

    \item
    Also of note is that our present implementation has yet to be optimized for run-time performance:
    we conducted our experiment on CPU,
    as our implementation is still missing batching over task functions.

    \item
    \citet{nsps} used a limit of 13 operations for their synthesized programs.
    In order to restrict ourselves to a finite subset of an otherwise potentially infinite search space,
    we apply such a limit during task function generation as well, in our case using a maximum of 3.
    For the sake of simplicity, we have opted to share this limit,
    specified during dataset generation, with our synthesizer as well.
    This disregards any synthesized programs of
    correct behavior exceeding our complexity limit.


    \item
    Our present implementation unfortunately still defers full type inference in favor of \emph{local types} based on our unrolled grammar,
    meaning it is unable to fill in further type variables based on information elsewhere in the program tree.

    \item
    The interpreter library we presently use for type inference may fail in the face of ambiguous type variables,
    somewhat limiting the potential effectiveness of our type-filter model.

    \item
    For run-time performance considerations,
    our dataset presently includes the input-output examples for each task function,
    in fact additionally shared across instantiated parameter types.
    A proper approach here would perhaps ensure different input-output pairs would be used across training epochs.
    This could be achieved by either generating a greater variety upfront,
    then sampling from these during training, or by generating these input-output pairs on the fly during training,
    incurring an additional run-time performance cost.

    \item
    Moreover, while fixing such a sample size satisfies the constraints of our LSTM used in sample conditioning,
    this is essentially an unfortunate compromise;
    if a function is significantly more general in its applicability,
    allowing for a wider variety of input types,
    then being forced to pick a fixed sample size means simpler programs with few types may potentially be learned more easily than programs with e.g. a greater number of type instances.
    It would likely be preferable to offset this sample size fixing by e.g. giving more weight to hard programs (or samples, for that matter).

\end{itemize}

\subsection{Topics for future research}

As neural synthesis methods aimed at programming by example in the functional
programming domain is a broad topic encompassing a variety of design decisions,
we have had to leave quite some questions unanswered.
We will give an overview in this section of some of the questions raised during our design process specifically.

    While we have focused on types as features here,
    after each synthesis step,
    we might also \emph{pre-compile} even partial programs such as to provide the synthesizer with immediate feedback on whether a program type-checks;
    as with types, we similarly hypothesize neural program synthesis methods can benefit from using compilation checks as additional features.

    However, this would require a \emph{weakly supervised} neural synthesizer,
    i.e. using \emph{reinforcement learning},
    whereas our present synthesizer is based on the simpler \emph{strongly supervised} setup.
    As such, this was unfortunately out of scope for our current paper.


    While we have looked into the added value of types as features,
    this gives rise to an additional question:
    what kind of operators are most conducive to benefiting from type info?
    While we have briefly conjectured this to involve having few yet generically applicable operators,
    this question has fallen out of scope for our paper,
    and remains a question for future research.



    As we touched upon in Section \ref{sec:grammar},
    the merits of our \emph{unrolled grammar} approach are still up for empirical evaluation.
    As our present dataset allowed us to settle for this approach,
    we decided to regard this question as out of scope for our present paper.




    For supervising with types, we have so far used the types we were able to infer from sample inputs and outputs.
    However, for each such different type,
    these make up type \emph{instances} of the task function's type signature.

    When synthesizing an \emph{identity} function,
    the sample mappings $0 \rightarrow 0$ and $false \rightarrow false$
    might give us the instantiated type signatures
    $Int \rightarrow Int$ or $Bool \rightarrow Bool$,
    yet the \emph{true} underlying type signature of this task function is $\forall a . a \rightarrow a$.

    One might then conclude that, in order to synthesize the intended function,
    knowing its true type signature would be more valuable than the sum of (knowing) its parts.
    This poses some challenges however:
        What can we learn about the true type signature from its instances?
        Is a given type variable shared between multiple parameters?
        That is, how do we know it should be $\forall a . a \rightarrow a$ over say $\forall a b . a \rightarrow b$?
        How much can we infer of type constraints on such type variables?
        That is, do we know the function should be as general as $\forall a . a \rightarrow a$, as opposed to say having some type constraint e.g. $\forall a . \text{Enum a} \Rightarrow a \rightarrow a$?

    Now, for either question, we could err in either direction, toward \emph{common denominator} upper and lower bounds.

    Suppose we cannot guarantee the type variable is shared across its two parameters,
    and we consider $\forall a b . a \rightarrow b$ the minimum requirement that a function type signature must satisfy.
    This might be the bare minimum requirement,
    but trying the more specific variants such as $\forall a . a \rightarrow a$,
    while potentially not \emph{guaranteed} to cover all cases,
    would nevertheless still be a useful type signature to consider during our synthesis,
    as its additional constraint may well make for some welcome search space reduction.

    Instead, we could also consider the case in which we don't need to infer the type signature from out input/output samples,
    but it is simply given to us as additional information for our (modified) PBE exercise.
    It would be interesting to investigate to what extent such true type signatures
    might aid synthesis, and furthermore, to what extent these may be inferred.
    However, given all these questions surrounding the use of such type signatures,
    we have opted to leave this as a topic for future research.

\subsection{Conclusion}

We presented a way to incorporate programming types into a neural program synthesis approach for programming by example.
We generated a dataset in the functional programming context,
and demonstrated type information to improve synthesis accuracy even given a comparable number of parameters.
Finally, we suggest a number of topics of interest for future research in type-driven neural programming by example.

\pagebreak

\section{Acknowledgements} 

First of all, I would like to thank my supervisor Emile van Krieken,
who has actively supported me with his broader knowledge on
program synthesis and AI whenever my own familiarity failed me.

Furthermore, I would like to thank my assessor Annette ten Teije for approving
my research direction, despite it falling outside the well-trodden
path for either artificial intelligence or computer science separately.

I additionally thank Emilio Parisotto, who helped answer my questions to better understand their neuro-symbolic program synthesis architecture.~\citep{nsps}

I must also thank my fellow students, discussions with whom helped me
tremendously throughout this program --- without them,
I probably wouldn't have made it this far.

On the implementation, I am further indebted to the HaskTorch team,
including Austin Huang, Junji Hashimoto, Torsten Scholak, Sam Stites, and Adam Paszke,
any of whom have not hesitated to advise me when I got stuck
on technical aspects including HaskTorch, Nix, Haskell, or machine learning.

I also owe gratitude to my friend Alejandra Ortiz, who was there for me when I needed it.

Finally, I would like to dedicate this work to Jaques Lagerweij,
whom I never got to meet.

\pagebreak

\nocite{*}
\bibliographystyle{plainnat}  
\bibliography{references}

\pagebreak

\appendix

\section{Hyperparameters} \label{sec:hpar}

\subsection{Hyperparameters used for dataset generation} \label{sec:gen-params}

In this section we will describe the hyperparameter values we have used in our dataset generation.

We generate types to substitute into \emph{type variables} using a maximum of only \emph{one} level of nesting,
i.e. allowing type \emph{list of booleans} though not type \emph{list of lists of booleans}.

For any parameter type containing type variables used in task functions,
we generate a maximum of $5$ type \emph{instances}, before deduplication.
Whereas \citet{nsps} generated $10$ inputs for each task function,
we instead generate up to $10$ for each \emph{type instance} of a task function, before deduplicating.

While they limited functions to a maximum of $13$ operations,
we instead limit ours to a maximum of $3$,
given that our current operator set is considerably bigger than those of their \emph{FlashFill} domain.

Numbers that we generate, all of them integers,
we limit to the range from $-20$ to $20$.
For characters we stick to the range of digits, i.e. from `0' to `9',
a decision made with the intent to let their characters overlap
with those of digits for the purpose of helping reduce characters used in the encoder,
in turn reducing the size of its one-hot embedding.
This arbitrary constraint serves no other purpose than to constrain required compute.

Data structures such as \emph{string}, \emph{list}, \emph{set}, and \emph{hashmap},
we each generate using lengths in the range from $0$ to $5$.
Of these, \emph{sets} might further deduplicate down,
as this structure only holds unique items.

Our dataset we split into training, validation and test sets
using a ratio of $35\%$, $35\%$, and $30\%$, respectively.
As \citep{nsps} we sample $1,000$ training programs from the total function space.

\subsection{Hyperparameters in our synthesizer} 

In this section we will describe the hyperparameter values we have during the training and evaluation of our synthesizers.

We use $3$ layers in our LSTMs, which are present in our sample encoder
(for both input and output), our type encoders (for rule expansions and holes),
as well as for sample conditioning and scoring in our \emph{R3NN}.
We do allow bias terms although the original paper did not show these in their formulas.
We train for a maximum of $1,000$ epochs.

Our \emph{encoders} process items (either input-output samples or types) using a batch size of $8$.
Our \emph{R3NN} must use a fixed number of embedded input-output pairs on the basis of its LSTM used for conditioning,
and as such we have fixed this to use samples of $8$ embedded input-output pairs.

As \citet{nsps}, for synthesizer evaluation we sample $100$ functions from the model for each task function,
determining success based on the best from this sample,
i.e. considering the synthesis a success if any of these pass our PBE task, exhibiting the desired behavior.

We evaluate performance on our validation set once after every $5$ epochs of training.
During evaluation we similarly check for convergence based on the loss,
averaging over windows of $2$ evaluations,
i.e. stop training if the validation loss over the past two evaluations has increased from the two before.


We arbitrarily limit synthesized functions to the same complexity limit of $6$ operators as used during generation of task functions.

We allow $32$ features in our symbol and expansion rule embeddings, i.e. $M$ in \citet{nsps}'s \emph{R3NN}.
We allow $32$ features per input or output per LSTM direction, i.e. $H$ in \citet{nsps}'s \emph{sample encoder}.

We clip gradients to stay within a range from $-1$ to $1$.

The learning rate for our Adam optimizer we search over
by a grid search using our vanilla NSPS model, considering
values of \num{1e-2}, \num{1e-3}, \num{1e-4}, and \num{1e-5}.
Of these, we settle on a learning rate of \num{1e-2}.

\section{Miscellaneous experiments}

Aside from our main experiment, we also tried a few other configurations for which we had not managed to obtain conclusive results.

\subsection{Type filter}

The first of these was the idea to combine a synthesizer with a compiler check to filter out any non-compiling programs.
While the downside to this would be that the synthesizer would be made dependent on this extra compiler check,
incurring a run-time penalty during synthesis, linear in the number of expansion rules provided,
the advantage to such a setup would be that the synthesizer would no longer need to learn to disregard non-compiling programs itself,
reducing synthesis to a ranking problem of the compiling (partial) candidate programs.
%
%
We achieve this by simply masking the predicted scores of uncompiling programs in our NSPS implementation
(before calculating actual probabilities by softmax) to have no probability, i.e. $p(e)=0.0$.

While we failed to obtain any significant improvement over the baseline model using this setup,
this result may well have related to our implementation.
We presently used the \verb|hint| Haskell library as our interpreter for type-checks,
which unfortunately yielded false positive compiler errors for types containing ambiguous type variables,
such as \verb|show undefined|, which the Haskell compiler would resolve to type \emph{string},
whereas the \verb|hint| library would complain that the \verb|undefined|
argument would prevent resolving \verb|show|'s type variable.~\citep{hintambiguous}

As this counter-factual signal would prevent this synthesizer from correctly synthesizing the affected programs,
the fact that it nevertheless performed on par with our baseline algorithm suggests this approach does in fact have potential.
While we might have addressed this flaw in our implementation by switching
from this interpreter library to using Haskell's compiler API directly,
due to time constraints this unfortunately fell out of scope for this thesis.

\subsection{Picking holes}

Although the topic of which hole to fill was not directly touched upon in \citet{nsps},
our baseline implementation had the synthesizer deterministically fill the first hole
(under any given order --- we used left-to-right).
Nevertheless, we did also wonder what the effect might be if we would allow filling any hole.

During training, we would then opt to randomly pick a hole to try and fill.
On evaluation, we would then look at the confidence scores for any hole expansions across holes,
sampling from this full matrix rather than just the vector slice corresponding to the first hole.
This allows the synthesizer to take into account the relative confidence of expansions for different holes,
enabling it to forego holes involving more uncertainty in favor of those it feels more confident about,
which may in turn provide additional information that may then reduce ambiguity for the remaining holes.%
~\footnote{
    An additional advantage of this would be it could more uniformly explore various partial program trees across synthesis steps.
    That said, uniform exploration there isn't necessarily the ideal situation ---
    one might for example imagine using weights to prioritize situations our synthesizer is less confident about.
}

Unfortunately, we obtained inconsistent results on this model versus our baselines across different experiment attempts,
originally getting the expected improvement, although in our final implementation we had not managed to reproduce this improvement.
We had to leave further analysis of these inconsistent results out of scope due to time constraints,
and as such feel hard-pressed to make definitive statements on the effectiveness of this approach.
Nevertheless, we consider this to be a topic of interest in AST-based neural program synthesis.

\end{document}

%% file: title-page-ai.tex
\begin{titlepage}

\newcommand{\HRule}{\rule{\linewidth}{0.5mm}} 
\center 
 

\includegraphics[width=\linewidth]{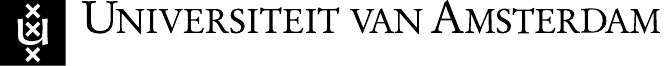}\\[2.5cm]
\textsc{\Large MSc Artificial Intelligence}\\[0.2cm]
\textsc{\Large Master Thesis}\\[0.5cm] 


\HRule \\[0.4cm]
{ \huge \bfseries
    Type-driven Neural Programming by Example \\
}
\HRule \\[0.5cm]
 

by\\[0.2cm]
\textsc{\Large
        Kiara Grouwstra
}\\[0.2cm]
    6195180
\\[1cm]


{\Large \today}\\[1cm]

    48
\\
    2019-2020
\\[1cm]

\begin{minipage}[t]{0.4\textwidth}
\begin{flushleft} \large
\emph{Supervisor:} \\
    MSc. Emile \textsc{van Krieken}
\end{flushleft}
\end{minipage}
~
\begin{minipage}[t]{0.4\textwidth}
\begin{flushright} \large
\emph{Assessor:} \\
    Dr. Annette \textsc{ten Teije}
\\
\end{flushright}
\end{minipage}\\[2cm]
~
\begin{minipage}[t]{0.4\textwidth}
\begin{flushleft} \large
\emph{} \\
\end{flushleft}
\end{minipage}
~
\begin{minipage}[t]{0.4\textwidth}
\begin{flushright} \large
\emph{Second Reader:} \\
    Dr. Clemens \textsc{Grelck}
\\
\end{flushright}
\end{minipage}\\[2cm]
    

\includegraphics[width=2.5cm]{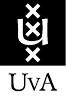}\\ 
\textsc{\large
        University of Amsterdam\\Informatics Institute
}\\[1.0cm]
 

\vfill 

\end{titlepage}